\documentclass[lettersize,journal]{IEEEtran}
\usepackage{amsmath,amsfonts}
\usepackage{algorithmic}
\usepackage{array}
\usepackage[caption=false,font=normalsize,labelfont=sf,textfont=sf]{subfig}
\usepackage{textcomp}
\usepackage{stfloats}
\usepackage{url}
\usepackage{verbatim}
\usepackage{graphicx}
\def\BibTeX{{\rm B\kern-.05em{\sc i\kern-.025em b}\kern-.08em
    T\kern-.1667em\lower.7ex\hbox{E}\kern-.125emX}}
\usepackage{balance}

\usepackage[colorlinks,urlcolor=blue,linkcolor=blue,citecolor=blue]{hyperref}
\usepackage{color}
\definecolor{Gray}{rgb}{0.572,0.572,0.572}
\usepackage{color,array}

\usepackage{algorithm}

\usepackage[caption=false,font=normalsize,labelfont=sf,textfont=sf]{subfig}

\usepackage{cite}
\usepackage{romannum}
\usepackage{diagbox}
\usepackage{makecell}
\usepackage{multirow}
\usepackage{ragged2e} 
\usepackage{float}
\usepackage{subcaption}
\usepackage{enumitem}
\usepackage{tabularray}
\usepackage{lscape}
\usepackage{multirow}
\usepackage{longtable}

\begin{document}

\title{Beyond Uncertainty Quantification:\\ Learning Uncertainty for Trust-Informed Neural Network Decisions –
A Case Study in COVID-19 Classification}

\author{Hassan Gharoun\textsuperscript{1}, Mohammad Sadegh Khorshidi\textsuperscript{1}, Fang Chen\textsuperscript{1}, and Amir H. Gandomi\textsuperscript{1,2,3}
\thanks{\textsuperscript{1}Faculty of Engineering \& IT, University of Technology Sydney, e-mails: Hassan.Gharoun@student.uts.edu.au, Mohammadsadegh.khorshidialikordi@student.uts.edu.au, Fang.Chen@uts.edu.au, Gandomi@uts.edu.au.}
\thanks{\textsuperscript{2}University Research and Innovation Center (EKIK), Óbuda University.}
\thanks{\textsuperscript{3}Corresponding author}}



\maketitle

\begin{abstract}
Reliable uncertainty quantification is critical in high-stakes applications, such as medical diagnosis, where confidently incorrect predictions can erode trust in automated decision-making systems. Traditional uncertainty quantification methods rely on a predefined confidence threshold to classify predictions as confident or uncertain. However, this approach assumes that predictions exceeding the threshold are trustworthy, while those below it are uncertain, without explicitly assessing the correctness of high-confidence predictions. As a result, confidently incorrect predictions may still occur, leading to misleading uncertainty assessments. To address this limitation, this study proposed an uncertainty-aware stacked neural network, which extends conventional uncertainty quantification by learning when predictions should be trusted. The framework consists of a two-tier model: the base model generates predictions with uncertainty estimates, while the meta-model learns to assign a trust flag, distinguishing confidently correct cases from those requiring expert review. The proposed approach is evaluated against the traditional threshold-based method across multiple confidence thresholds and pre-trained architectures using the COVIDx CXR-4 dataset. Results demonstrate that the proposed framework significantly reduces confidently incorrect predictions, offering a more trustworthy and efficient decision-support system for high-stakes domains.
\end{abstract}


\section{Introduction}

\IEEEPARstart{T}{he} COVID-19 pandemic affected numerous countries, leading to millions of cases and fatalities worldwide. Detecting and diagnosing COVID-19 at an early stage proved to be crucial in controlling its spread \cite{jin2020development}. Rapid diagnosis facilitated timely medical intervention and recovery. Alongside polymerase chain reaction (PCR) tests, chest radiography (X-rays) and computed tomography (CT) scans were employed as key diagnostic tools. Due to the limited availability of PCR tests in many regions, medical imaging emerged as a primary method for diagnosing COVID-19 \cite{jin2020development}. As a standard procedure, these images required manual analysis by clinical experts. However, the ongoing shortage of healthcare professionals, especially in developing nations and smaller hospitals, made quick diagnosis challenging and exacerbated the workload on existing experts. At this moment, the advantages of AI and machine-learning (ML) became more obvious in healthcare. 

However, it appears that no AI-based healthcare tools have been officially incorporated into clinical guidelines as a standard part of medical practice \cite{burger2024unmet}. A critical factor that hinders AI acceptance and adoption among both healthcare professionals and patients is the lack of trust in the decisions made by AI/ML systems \cite{cheng2022promoting}.

Numerous studies have investigated the underlying factors contributing to the persistent challenge of establishing trust in AI within the healthcare domain, and several key factors have been identified \cite{cheng2022promoting}. One major factor contributing to the lack of trust in AI systems is their inconsistent performance across different clinical settings \cite{fehr2024trustworthy}. One factor that can cause variations or inconsistencies in a model’s performance across different conditions is the model’s uncertainty. 

Model uncertainty can be defined as the variability in the model’s confidence estimations. Ideally, a model’s confidence score—representing the probability assigned to the predicted class—should align with the true likelihood of correctness. For instance, if a model assigns a 90\% confidence score to a prediction, then, across many similar instances, approximately 90\% of those predictions should be correct. However, confidence scores alone do not guarantee reliability, as their consistency can vary across different cases.

Uncertainty mainly originates from two sources \cite{abdar2021review}: (\Romannum{1}) data uncertainty, which is due to elements such as noise, complexity, and limited knowledge about environmental conditions (aleatoric uncertainty), and (\Romannum{2}) parametric uncertainty, which occurs when the model is inadequate because of imprecise understanding of its components (epistemic uncertainty).

The presence of various sources of uncertainty makes it crucial to determine how much a model’s confidence can be trusted. To this end, uncertainty quantification techniques have been developed to measure and communicate the reliability of a model’s predictions. Typically, uncertainty is quantified and compared against a predefined threshold: values above this threshold are flagged as \textit{'uncertain'}, whereas those below are deemed \textit{'confident.'}

Ideally, in a well-calibrated model, correct predictions would exhibit lower uncertainty (falling below the threshold), while incorrect predictions would exhibit higher uncertainty (exceeding the threshold). However, in practice, some incorrect predictions can still fall below this threshold, resulting in confidently incorrect outcomes. This conventional approach, where confidently incorrect predictions are not adequately addressed, exacerbates the erosion of trust in ML and AI, particularly in high-stakes fields such as medical diagnosis.

\textbf{Contribution.} In a production environment, simply reporting the predicted class alongside its quantified uncertainty—whether above or below a predefined threshold—may not effectively prevent confidently incorrect predictions, as discussed earlier. Accordingly, this study target this limitation and proposes a framework for determining whether a prediction should be trusted.

The proposed framework consists of a two-tier architecture with a dual-output system: 
\begin{itemize}
    \item Tier 1: Generates the predicted outcome along with its associated uncertainty estimates.
    \item Tier 2: Produces a binary flag indicating whether the prediction is confidently correct and can therefore be trusted.
\end{itemize}

The second-tier model is trained using the original input features, along with the quantified uncertainty, and ground truth data (available during training). Here, objective is to learn patterns and relationships among the input features, predictions, and uncertainty estimates to determine whether a prediction is confidently correct and can be trusted. At test time, when the ground truth is unknown, the trained second-tier model leverages the predictions and their quantified uncertainty to assess whether the predictions should be trusted assisting the operator or end-user.
 
This second output enables the model to flag predictions with a "do not trust me" alert, prompting human experts (such as physicians in medical applications) to re-evaluate them. Thus, the model not only enhances user experience but also fosters greater adoption by encouraging cautious and trust-informed decision-making.

The rest of the paper is organized as follows: Section \ref{section:Related works} reviews relevant previous research and outlines the contributions of this study. Section \ref{section:proposed method} provides a detailed exploration of the proposed algorithm. Sections \ref{Section:Dataset} and \ref{Section:Setup} explain the dataset used and describe the experimental framework respectively. Insights and analysis from the experiments are presented in Section \ref{section:Discussion and Results}. Finally, Section \ref{section:Conclusion} offers concluding remarks and suggests directions for future research.

\section{Background} \label{section:Related works}
ML including neural networks (NNs) have reached a peak in accuracy. However, 
confidence estimations (represented by prediction probabilities), are often used to interpret a model's accuracy. However, these estimations can be unreliable and prone to variation, leading to doubts about how well they represent true prediction correctness.

To evaluate variability in confidence estimations, it is crucial to recognize that most conventional ML models, including NNs, typically generate deterministic predictions by producing a single output, or point estimate, for a given input \cite{aseeri2021uncertainty}. This deterministic approach fails to capture the variability (or in other words uncertainty) in the model's predictions. 

To illustrate this, consider a neural network \( f \) parameterized by \( \mathbf{w} \), which maps an input \( \mathbf{x} \in \mathcal{X} \) to an output \( y \in \mathcal{Y} \). The target output is obtained by optimizing the parameters of the network, such that \( y = f_{\mathbf{w}}(\mathbf{x}) \). Rather than relying on point estimation for 
\( \mathbf{w} \), assigning a probability distribution over the model parameters enables the derivation of a probability distribution for predictions, allowing for the quantifying uncertainty regarding the model's knowledge (referred to as epistemic uncertainty). Bayesian inference methods - including Markov chain Monte Carlo (MCMC) \cite{gamerman2006markov}, Variational inference (VI) \cite{graves2011practical}, Monte Carlo dropout (MCD) \cite{gal2016dropout}, Variational Autoencoders (VAE) \cite{kingma2013auto}, Bayes By Backprop (BBB) \cite{fortunato2017bayesian} - are commonly employed to estimate the posterior distribution of the model parameters to achieve this. In addition to Bayesian techniques, ensemble learning is another method frequently used to quantify uncertainty. In a typical ensemble, each model independently predicts the output for a given input. When these models are diverse—built with different architectures, parameters, or trained on various data subsets—they produce probabilistic predictions instead of single-point estimates. For further details on uncertainty quantification techniques, interested readers can refer to \cite{abdar2021review}.

By leveraging the aforementioned uncertainty quantification (UQ) methods, there has been a growing interest in the development of uncertainty-aware ML models, particularly NNs. These research efforts can be categorized into two main categories:

The first category of studies emphasizes quantifying uncertainty to enhance decision-making by communicating prediction uncertainties. Typically, these studies generate predictions accompanied by uncertainty estimates using one of the aforementioned UQ methods. Predictions with the highest uncertainty are flagged as potentially inaccurate, enabling more informed and cautious decision-making. Among these studies, MCD is widely employed in healthcare to quantify uncertainty and improve decision-making. For instance, in cardiac arrhythmia detection, gated recurrent neural networks (GRUs) with MCD provide well-calibrated uncertainty estimates, crucial for clinical confidence \cite{aseeri2021uncertainty}. Similarly, deep learning models with MCD are utilized for stroke outcome prediction, helping to identify high-risk predictions that necessitate further human evaluation\cite{martin2024uncertainty}. Another application involves the multi-level context and uncertainty aware (MCUa) model for breast histology. The model uses context-aware networks to learn spatial dependencies among image patches and applies MCD to measure confidence levels based on the standard deviation of multiple predictions. Lower standard deviations are interpreted as confident. \cite{senousy2021mcua}. Furthermore, MCD enhances colorectal polyp classification by utilizing predictive variance and entropy for uncertainty measurement, along with temperature scaling for confidence calibration \cite{carneiro2020deep}. Application of MCD is not limited to health care and in the domain of credit card fraud detection enhancing the clarity of the system's reliability in the detection process \cite{habibpour2023uncertainty}.

While MCD is a popular choice in healthcare, Bayesian deep learning techniques have found application across a wider range of domains, providing robust uncertainty quantification. In engineering applications, Bayesian neural networks (BNNs) and stochastic variational Gaussian processes (SVGPs) are used for building energy modeling, providing predictions with confidence intervals that optimize resource allocation and enhance model robustness\cite{westermann2021using}. In nuclear power plants, Bayesian models estimate predictive uncertainty to enhance decision-making and risk management in health monitoring systems\cite{yao2024uncertainty}. Another study leverages Bayesian networks in agribusiness risk assessment to quantify uncertainty and improve out-of-domain calibration, aiding financial decision-making\cite{teixeira2023bayesian}. Furthermore, in transformer diagnostics, Gaussian Bayesian networks (GBNs) are combined with black-box classifiers to quantify uncertainty, leveraging the strengths of different models to improve diagnostic accuracy through probabilistic predictions\cite{aizpurua2018uncertainty}.

In addition to MCD and Bayesian approaches, ensemble learning methods are prominently used for uncertainty quantification, particularly in high-stakes environments where reliability is paramount. A framework for COVID-19 diagnosis employs pre-trained convolutional neural networks (CNNs) like VGG16 and ResNet50, extracting features from chest X-rays and CT images while estimating epistemic uncertainty through model ensembles to ensure higher accuracy and reliability\cite{shamsi2021uncertainty}. In defect detection for casting products, transfer learning with CNNs and deep learning ensembles is used to estimate epistemic uncertainty, improving model trustworthiness by identifying poorly trained input regions\cite{habibpour2021uncertainty}. In food recognition tasks, epistemic uncertainty guides the selection of ensemble models, enhancing accuracy and robustness by choosing diverse models with low mean uncertainty\cite{aguilar2022uncertainty}. 

The second category of studies goes beyond merely quantifying uncertainty, incorporating it directly into the training process to enhance the model's confidence. Typically, these approaches introduce new loss functions that are designed around uncertainty estimations, allowing the model to better account for uncertainty during learning. Most studies extend the standard cross-entropy loss by combining it with auxiliary terms—such as expected calibration error (ECE), predictive entropy (PE), or Kullback-Leibler (KL) divergence—to better align predicted uncertainty with actual outcomes \cite{tabarisaadi2022optimized,shamsi2023novel,shamsi2021uncertainty_loss,li2022ultra}. Additionally, some works introduce novel, innovative loss functions (e.g., paired confidence loss by \cite{dawood2023uncertainty}, accuracy versus uncertainty (AvUC) lossand maximum mean calibration error (MMCE) by \cite{dawood2024_improving}) that specifically address overconfident mistakes and enhance model calibration.

The first category of research shows that uncertain predictions can be effectively identified, while the second demonstrates that leveraging quantified uncertainty as a loss term improves both accuracy and calibration. However, none of these approaches fully eliminates confidently incorrect predictions, since most simply apply a fixed uncertainty threshold—a practice inherently insufficient for reliably detecting such cases. However, this question remains open: \textit{How can confidently incorrect predictions be averted while still preserving the benefits of uncertainty quantification and maintaining overall accuracy?} This study aims to address this question by proposing a stacked neural networks architecture designed to identify predictions that the model is highly confident in being correct.

\section{Methodology}\label{section:proposed method}
Stacked generalization, commonly known as stacking, is a robust ML technique that involves two distinct levels of models: the base models (level-0) and the meta-model (level-1). In the first level, various algorithms, which can be either heterogeneous (different types of models) or homogeneous (same type of models) \cite{park2022stacking}, are trained on the original training dataset. These base models generate predictions which are then used as input features for the meta-model. The meta-model's purpose is to learn the optimal way to combine these base models' predictions to achieve the best possible performance \cite{chatzimparmpas2021empirical}. Additionally, the meta-model can also incorporate the original input features from the training data alongside the base model outputs to enhance its learning process.

This study draws upon the concept of traditional stacking models to propose a new architecture that, while resembling stacking, serves a different purpose: trust-informed prediction. Here, the meta-model aims to learn the relationship between the base-model predictions and their associated uncertainties. Consequently, the output of the meta-model is a flag that indicates whether the model's prediction is trustworthy or not. In this context, 'trustworthy' specifically refers to predictions that are both correct and confident, as only such predictions are considered reliable for end-user decision-making. To align with this new objective, the architecture of the proposed method is described in detail in the following sections.

\subsection{Uncertainty-Aware Stacked Neural Networks}
\begin{figure}[]
  \centering
  \captionsetup[subfloat]{font=tiny}
  {\includegraphics[width=\columnwidth]{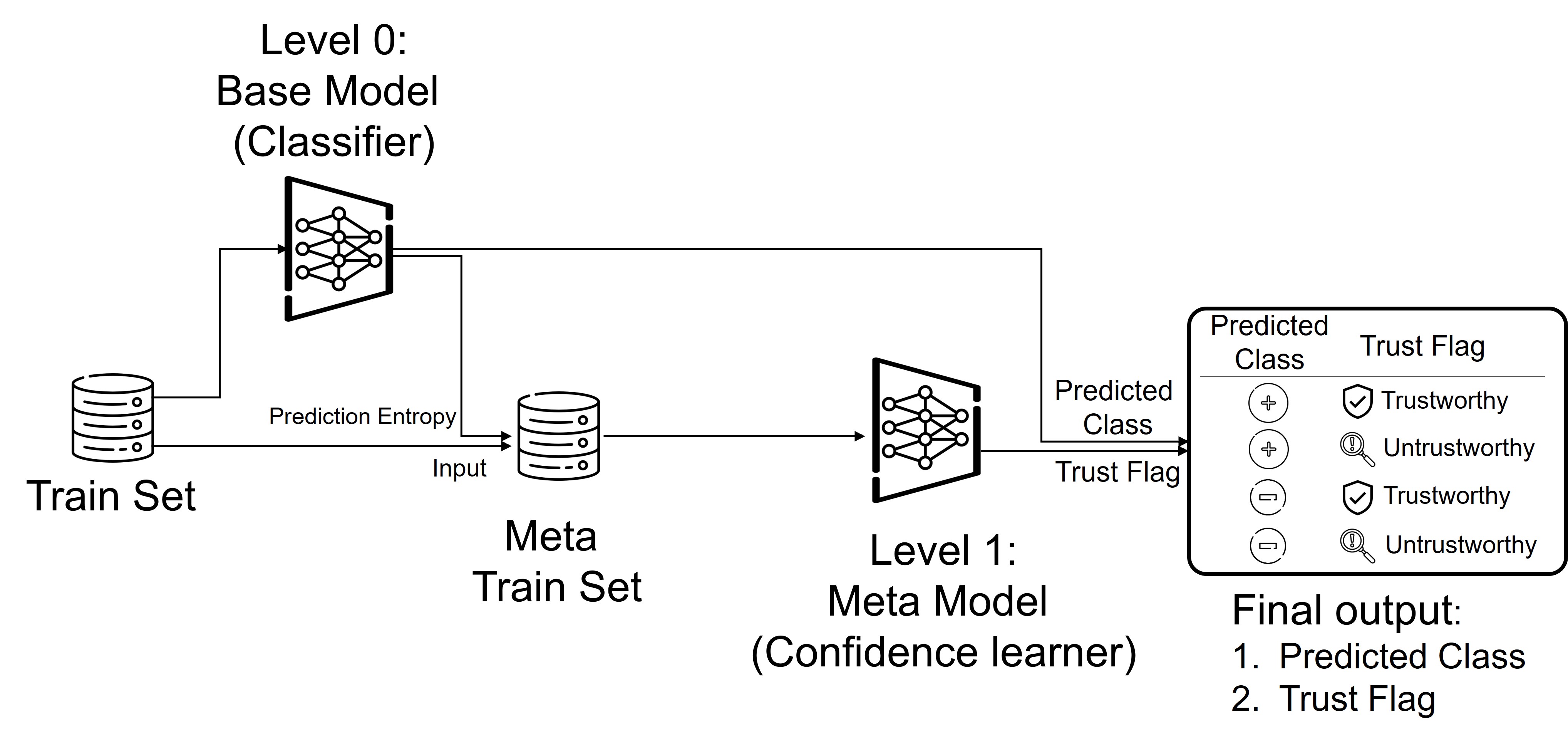}}\par
  \caption{Uncertainty-aware Stacked Neural Networks (U-SNN) Schema.}
  \label{fig:Confidence_aware_model_schema}
\end{figure}

Figure \ref{fig:Confidence_aware_model_schema} illustrates the overall flow of the uncertainty-aware stacked neural networks (U-SNN) framework, which is architected with two integrated layers: the Level 0 base model as the initial predictor and the Level 1 meta-model as the trust evaluator. This study applies the U-SNN framework to the classification task, beginning with  a dataset \( D = \{(x_1, y_1), (x_2, y_2), \ldots, (x_N, y_N)\} \), where \( x_i \) and \( y_i \) represent the $i^{th}$ observation vector and its corresponding label in a \(d\)-dimensional feature space. The initial step involves splitting the dataset into training and testing subsets, denoted as \( D_{\text{train}} \) and \( D_{\text{test}} \), respectively. In this study, \( D_{\text{train}} \) is exclusively used for training both the base model and the meta-model, while \( D_{\text{test}} \) is reserved solely for evaluating the performance of the proposed method. This approach ensures that the base and meta-models do not have access to or influence from the test dataset, thereby providing an unbiased assessment of the proposed method.

The base model classifier generates class predictions, with the associated uncertainty estimations used to construct a new training set for the meta-model. For clarity, this second training set is referred to as the meta-train set. The uncertainty estimations of the base model in this study is calculated by using MCD technique. Detailed description of MCD technique is provided in the subsequent section \ref{SubSec:UQ}. 

Following the training of the base model, the meta-train set is constructed by combining the original input features \( X_i \) with the estimated uncertainty \( e_i \) as an additional independent variable.

The target variable of the meta-train set is a binary label \( z_i \) derived as follows:

\begin{equation}
z_i = 
\begin{cases} 
1 & \text{if } \hat{y}_i = y_i \text{ and } e_i \leq \tau \\
0 & \text{otherwise } 
\end{cases}
\label{Eq:Meta_train_set_label}
\end{equation}

Where:
\begin{itemize}
    \item \( \hat{y}_i \): Predicted label for the $i^{th}$ instance
    \item \( y_i \): Ground truth for the $i^{th}$ instance
    \item \( e_i \): estimated prediction's uncertainty for the $i^{th}$ instance
    \item \( \tau \): Confidence threshold.
    \item \( z_i \): Binary label for the $i^{th}$ instance, with 1 indicating "Trustworthy" and 0 indicating "Untrustworthy".
\end{itemize}

Eq. \ref{Eq:Meta_train_set_label} reflects the primary objective of the meta-model, which is to identify instances where the model's predictions are both correct and confident ($z_i=1$). This conservative approach ensures that only predictions with high confidence and accuracy are trusted, while confidently incorrect predictions and uncertain predictions—whether correct or incorrect—are flagged as untrustworthy ($z_i=0$) and marked for further investigation. 

It is worth noting that ground truth labels are only available during training, whereas they remain unknown during testing and in real-world deployment. Consequently, the meta-model is designed to scrutinize any confident prediction from the base model, determining whether it might resemble a confidently incorrect instance and should be avoided. This mechanism mitigates the risk of trusting highly confident yet erroneous outputs.

In this study, prediction entropy (PE) is chosen as the sole metric indicator of prediction uncertainty. Therefore, in the Eq. \ref{Eq:Meta_train_set_label}, the estimated uncertainty $e_i$ denotes the PE. PE ranges between 0 and 1, with values closer to zero indicating lower entropy, and thus lower uncertainty and higher confidence, while values closer to 1 indicate higher entropy, greater uncertainty, and lower confidence. Consequently, a lower confidence threshold is more strict, as it allows only correct predictions with a PE lower than the $\tau$ to be flagged as trustworthy.

The following outlines the calculation of PE using the MCD. 

% The goal of employing these three distinct UQ techniques is to analyze how the uncertainty estimates obtained from different methods influence the model's ability to distinguish between confidently correct and the rest.

\subsection{Uncertainty Quantification with Monte-Carlo Dropout (MCD)} \label{SubSec:UQ}
% \subsubsection{}

Building on the concepts of UQ outlined in the section \ref{section:Related works}, MCD is implemented to approximate Bayesian inference in deep neural networks. Dropout originally designed as a regularization technique to prevent overfitting by randomly deactivating a subset of neurons during training.

This process can be adapted for uncertainty estimation by applying dropout during the inference stage. By performing multiple forward passes through the network with dropout enabled during inference, each pass results in slightly different predictions due to the random dropout of neurons, thereby producing a distribution of outputs for a given input. Each neuron in the network effectively samples from a Bernoulli distribution, and the collection of predictions across multiple stochastic forward passes serves as a Monte Carlo approximation of the posterior distribution.

Here, PE as an uncertainty evaluation metric is calculated as follows \ref{Eq:PE}:

\begin{equation}
    PE(\mathbf{x}) = - \sum_{c=1}^{C} \mu_{\text{pred}} (\mathbf{x}, c) \log[\mu_{\text{pred}} (\mathbf{x}, c)]
    \label{Eq:PE}
\end{equation}

Eq. \ref{Eq:PE} represents the prediction entropy $PE(\mathbf{x})$, calculated over $C$ classes, where $\mu_{\text{pred}}(\mathbf{x}, c)$ denotes the mean predicted probability of class $c$ for the input $\mathbf{x}$ calculated as Eq. \ref{Eq:Mean_pred_MCD}:

\begin{equation}
    \mu_{\text{pred}}(x, c) = \frac{1}{M} \sum_{m=1}^{M} p(y = c \mid x, \omega_m)
    \label{Eq:Mean_pred_MCD}
\end{equation}

where $p(y = c \mid x, \omega_m)$ denotes the probability that the input $x$ is assigned to class $c$, as determined by the softmax function, using the set of parameters $\omega_m$ from the $m^{\text{th}}$ iteration of the model, and $M$ signifies the count of such iterations.

\subsection{Evaluation Metrics} \label{Section:Evluation}
In the traditional approach, predictions are categorized as 'confident' or 'uncertain' based on whether their PE falls below or above a predefined threshold. To ease reference, this approach is hereafter referred to as the \textit{threshold-based} method.  

To evaluate and compare the proposed U-SNN framework with the threshold-based approach, a common set of evaluation metrics was introduced. These metrics are designed to assess two critical aspects: (a) the effectiveness of each approach in minimizing confidently incorrect predictions and (b) the efficiency in optimizing the referral process for uncertain cases requiring further review.

Although the same metrics are applied, they are calculated differently for the threshold-based method and the U-SNN method, reflecting the distinct architectures and mechanisms used to manage uncertainty.

A key clarification is necessary regarding terminological consistency. The U-SNN method explicitly classifies predictions as trustworthy or untrustworthy. According to the definition provided in Eq. ~\ref{Eq:Meta_train_set_label}, a trustworthy prediction corresponds to a confidently correct prediction in the threshold-based method. Conversely, an erroneously flagged trustworthy prediction (i.e., a false trustworthy outcome) aligns with a confidently incorrect prediction in the threshold-based approach.

Considering this alignment, the terms confident, uncertain, correct, and incorrect are used consistently throughout the evaluation for ease of comparison. Although U-SNN does not explicitly label predictions as confident or uncertain, its outputs can be conceptually mapped to these terms. This approach ensures terminological consistency while maintaining clarity regarding how each method generates its outputs.

Building upon these clarifications, the following metrics are introduced:
\begin{itemize}
    \item Certainty Rate (CR): Measures how often the model is confident.
    \item False Certainty Rate (FCR): Captures how often the model is confidently incorrect among all predictions. 
    \item Confidence Error (CE): Quantifies among confident predictions, how often the model is incorrect. 
    \item Uncertainty Rate (UR): Denotes how often the model flags predictions for review.
    \item Redundant Referral (RR): Reflects among flagged cases, how many were unnecessary.
\end{itemize}

\noindent In the following subsections, the calculation of each of the above metrics is defined for both the threshold-based method and the proposed U-SNN approach, considering the distinct mechanisms of each method.

\subsubsection{Metric Formulation for the Threshold-Based Method}
In the traditional threshold-based method, combining predictions' correctness (correct or incorrect) with their uncertainty status (confident or uncertain) yields four distinct outcome categories.\cite{habibpour2023uncertainty}:

\begin{itemize}
    \item True Certainty (TC): Predictions correctly classified by the model confidently.
    \item False Certainty (FC): Predictions incorrectly classified by the model but labeled as confident.
    \item True Uncertainty (TU): Incorrect predictions correctly identified by the model as uncertain, appropriately flagged for further review.
    \item False Uncertainty (FU): Predictions correctly classified but flagged as uncertain, resulting in redundant reviews.
\end{itemize}

Considering the above outcome categories, the metrics described in the previous section are formulated as follows:

\begin{itemize}
    \item CR: Measures the proportion of predictions labeled as confident out of all predictions denoted by Eq. \ref{eq:CR}. 

    \begin{equation}
        CR= \frac{TC+FC}{TC+FC+TU+FU}
        \label{eq:CR}
    \end{equation}

    \item FCR: Measures the proportion of confidently incorrect predictions across all predictions represented by Eq. \ref{eq:FCR}.

    \begin{equation}
        FCR= \frac{FC}{TC+FC+TU+FU}
        \label{eq:FCR}
    \end{equation}

    \item CE: quantifies the error rate within predictions labeled as confident presented by Eq. \ref{eq:CE}.

    \begin{equation}
        CE= \frac{FC}{TC+FC}
        \label{eq:CE}
    \end{equation}

    \item UR: Reflects the fraction of samples flagged as uncertain and referred for additional review denoted by Eq. \ref{eq:UR}.

    \begin{equation}
        UR= \frac{TU+FU}{TC+FC+TU+FU}
        \label{eq:UR}
    \end{equation}

    \item RR: Represents the proportion of correct predictions unnecessarily flagged as uncertain, relative to all uncertain predictions formulated by Eq. \ref{eq:RR}.

    \begin{equation}
        RR= \frac{FU}{TU+FU}
        \label{eq:RR}
    \end{equation}
    
\end{itemize}

\subsubsection{Metric Formulation for the U-SNN Approach}
The proposed U-SNN method employs a dedicated meta-model trained specifically to discern whether a prediction from the base model is trustworthy or requires additional review. In a manner analogous to the traditional confusion matrix, the base model’s predictions are compared with the ground truth labels, resulting in four categories: true positive (TP), false positive (FP), true negative (TN), and false negative (FN). Concurrently, the meta-model’s predicted confidence labels are compared with the ground truth trustworthy labels, yielding four additional categories: true trustworthy (TT), false trustworthy (FT), true untrustworthy (TU), and false untrustworthy (FU). By integrating these correctness and confidence classifications, 16 distinct outcomes are generated, as illustrated in Table \ref{tab:New_Confusion_Matrix}. For ease of reference, the proposed confusion matrix is called the trust-informed confusion matrix.

\begin{table}[H]
\centering
\caption{Trust-informed Confusion Matrix}
\label{tab:New_Confusion_Matrix}
\resizebox{\columnwidth}{!}{%
\begin{tabular}{ccl|cccc|}
\cline{4-7}
\multicolumn{1}{l}{} & \multicolumn{1}{l}{} &  & \multicolumn{4}{c|}{\rule{0pt}{2.5ex} U-SNN Output} \\ \cline{4-7} 
\multicolumn{1}{l}{} & \multicolumn{1}{l}{} &  & \multicolumn{2}{c|}{\rule{0pt}{2.5ex} Positive} & \multicolumn{2}{c|}{\rule{0pt}{2.5ex} Negative} \\ \cline{4-7} 
\multicolumn{1}{l}{} & \multicolumn{1}{l}{} &  & \multicolumn{1}{c|}{\rule{0pt}{2.5ex} Trustworthy} & \multicolumn{1}{c|}{\rule{0pt}{2.5ex} Untrustworthy} & \multicolumn{1}{c|}{\rule{0pt}{2.5ex} Trustworthy} & \rule{0pt}{2.5ex} Untrustworthy \\ \hline
\multicolumn{1}{|c|}{\multirow{4}{*}{\rule{0pt}{2.5ex} \makecell{Ground\\Truth}}} & \multicolumn{1}{c|}{\multirow{2}{*}{\rule{0pt}{2.5ex} Positive}} & \rule{0pt}{2.5ex} Trustworthy & \multicolumn{1}{c|}{\rule{0pt}{2.5ex} TPTT} & \multicolumn{1}{c|}{\rule{0pt}{2.5ex} TPFU} & \multicolumn{1}{c|}{\rule{0pt}{2.5ex} FNTT} & \rule{0pt}{2.5ex} FNFU \\ \cline{3-7} 
\multicolumn{1}{|c|}{} & \multicolumn{1}{c|}{} & \rule{0pt}{2.5ex} Untrustworthy & \multicolumn{1}{c|}{\rule{0pt}{2.5ex} TPFT} & \multicolumn{1}{c|}{\rule{0pt}{2.5ex} TPTU} & \multicolumn{1}{c|}{\rule{0pt}{2.5ex} FNFT} & \rule{0pt}{2.5ex} FNTU \\ \cline{2-7} 
\multicolumn{1}{|c|}{} & \multicolumn{1}{c|}{\multirow{2}{*}{\rule{0pt}{2.5ex} Negative}} & \rule{0pt}{2.5ex} Trustworthy & \multicolumn{1}{c|}{\rule{0pt}{2.5ex} FPTT} & \multicolumn{1}{c|}{\rule{0pt}{2.5ex} FPFU} & \multicolumn{1}{c|}{\rule{0pt}{2.5ex} TNTT} & \rule{0pt}{2.5ex} TNFU \\ \cline{3-7} 
\multicolumn{1}{|c|}{} & \multicolumn{1}{c|}{} & \rule{0pt}{2.5ex} Untrustworthy & \multicolumn{1}{c|}{\rule{0pt}{2.5ex} FPFT} & \multicolumn{1}{c|}{\rule{0pt}{2.5ex} FPTU} & \multicolumn{1}{c|}{\rule{0pt}{2.5ex} TNFT} & \rule{0pt}{2.5ex} TNTU \\ \hline
\end{tabular}%
}
\end{table}

In the proposed trust-informed confusion matrix, four combinations—FNTT, FPTT, FNFU, and FPFU—never occur. This exclusion is rooted in the definition of the meta-train set target variable as described in Eq. \ref{Eq:Meta_train_set_label}. As per this definition, incorrect predictions are never classified as trustworthy; hence, false positives and false negatives cannot be designated as true trustworthy (TT). Furthermore, since incorrect predictions cannot qualify as true trustworthy, they are also precluded from being classified as false untrustworthy (FU). 

From this expanded structure, the evaluation metrics are reformulated as follows:

\begin{itemize}
    \item CR: Calculated by measuring the proportion of predictions classified as trustworthy, irrespective of correctness, denoted by Eq. \ref{Eq:CR_v2}. Importantly, all trustworthy predictions—regardless of their correctness—are conceptually equivalent to confident predictions in the threshold-based method.

    \begin{equation}
        CR = \frac{TTP}{Total\ Outcomes}
        \label{Eq:CR_v2}
    \end{equation}
    where TTP stands as total trustworthy predictions defined by Eq. \ref{Eq:TTP}:
    \begin{align}
        TTP = &\ TPTT + FNTT + FPTT + TNTT + \notag \\
              &\ TPFT + FNFT + FPFT + TNFT
        \label{Eq:TTP}
    \end{align}
    
    \item FCR: Measured by the proportion of false trustworthy predictions relative to all predictions, presented by Eq. \ref{Eq:FRC_v2}. Notably, false trustworthy predictions are conceptually equivalent to false confident outcomes in the context of the threshold-based method.

    \begin{equation}
         FCR = \frac{FPFT + FNFT}{Total\ Outcomes}
        \label{Eq:FRC_v2}
    \end{equation}

    \item CE: Quantified by the proportion of false trustworthy predictions among all trusted predictions, as calculated by Eq. \ref{Eq:CE_v2}. Importantly, the proportion of false trustworthy predictions among all trusted predictions conceptually corresponds to the proportion of false certain predictions among all confident predictions in the context of the threshold-based method.

    \begin{equation}
        CE = \frac{FPFT + FNFT}{TTP}
        \label{Eq:CE_v2}
    \end{equation}
    
    \item UR: Calculated by the proportion of predictions classified as untrustworthy (flagged for review) relative to all predictions, as defined in Eq. \ref{Eq:UR_v2}. 

    \begin{equation}
       UR = \frac{TUP}{Total\ Outcomes}
       \label{Eq:UR_v2}
    \end{equation}
    where TUP is defined as total untrustworthy predictions as Eq. \ref{Eq:TUP}:
    \begin{align}
    TUP = &\ TPFU + TPTU + FNFU + FNTU + \notag \\
          &\ FPFU + FPTU + TNFU + TNTU 
    \label{Eq:TUP}
    \end{align}

    \item RR: Quantifies the proportion of flagged predictions for review that were unnecessary, as calculated by Eq \ref{Eq:RR_v2}. In the U-SNN method, redundant referrals arise when predictions correctly classified by the base model are mistakenly flagged as untrustworthy by the meta-model. These include correctly predicted outcomes incorrectly labeled as untrustworthy (TPFU, TNFU) and correctly predicted outcomes flagged as uncertain (TPTU, TNTU). Importantly, TPTU and TNTU are conceptually equivalent to false uncertainty (FU) in the context of the threshold-based method, where correct predictions are unnecessarily flagged for review.

    \begin{equation}
        RR = \frac{TPFU + TNFU + TPTU + TNTU}{TUP}
        \label{Eq:RR_v2}
    \end{equation}

\end{itemize}

\section{Data set} \label{Section:Dataset}
\begin{figure}[]
  \centering
  \captionsetup[subfloat]{font=tiny}
  \subfloat[][Covid 19 positive samples]{\includegraphics[width=\columnwidth]{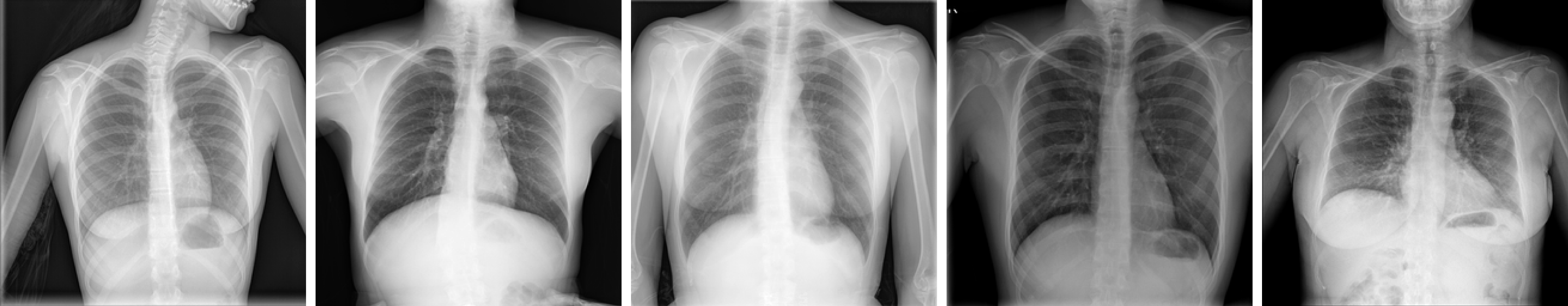}}\par
  \subfloat[][Covid 19 negative samples]{\includegraphics[width=\columnwidth]{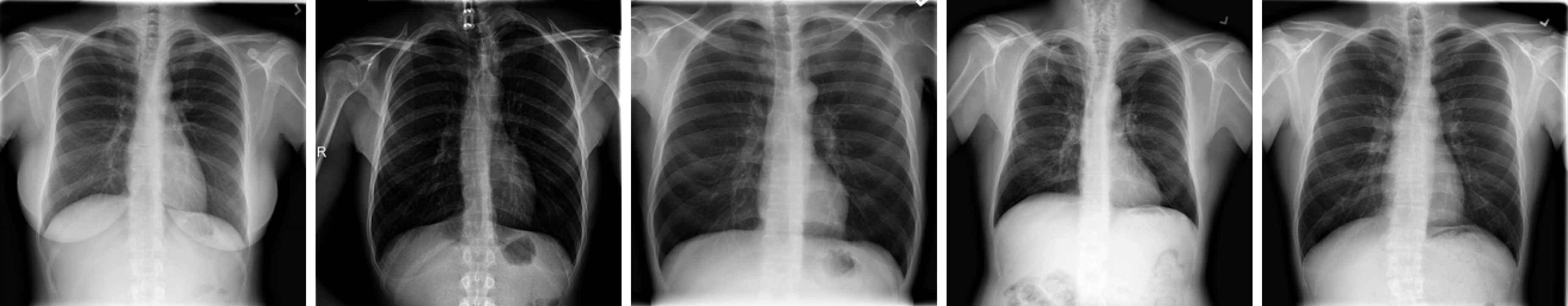}}\par
  \caption{Examples images from COVIDx CXR-4 dataset.}
  \label{fig:Sample_images_of_dataset}
\end{figure}

The dataset utilized in the study is COVIDx CXR-4 \cite{wu2023covidx}, an expanded multi-institutional open-source benchmark dataset specifically designed for chest X-ray image-based computer-aided COVID-19 diagnostics. This dataset significantly expands upon its predecessors, the COVIDx CXR datasets, by increasing the total patient cohort size to 84,818 images from 45,342 patients across multiple institutions. The age distribution of the patients ranges widely, though there is a notable bias, with over half of the patients being between 18 and 59 years old. Additionally, the dataset maintains a nearly equal gender distribution and varied imaging views. The COVIDx CXR-4 dataset includes two main classes: positive COVID-19 cases and negative cases. Among the 84,818 images, 65,681 are positive COVID-19 cases, while 19,137 are negative cases, reflecting a significant class imbalance. This dataset is publicly available \cite{kaggleCOVIDxCXR4}. Figure \ref{fig:Sample_images_of_dataset} shows samples of the COVIDx CXR-4 dataset. 

\section{Experimental Design} \label{Section:Setup}
\subsection{Transfer Learning as Image Embedding Generator}
This study adopted the transfer learning (TL) approach to prepare the dataset for the proposed method. Here, the primary goal of TL was to transfer knowledge by using pre-trained models as an embedding function (or, more simply, feature extractors). These models, initially trained and optimized on extensive datasets such as ImageNet, had their final fully-connected layers removed, and the remaining network layers were frozen. This repurposed the networks as fixed feature extractors for the COVIDx CXR-4 dataset, effectively harnessing their pre-established computational intelligence for new data applications.

In this study, three different pre-trained models were utilized: EfficientNetB0 \cite{tan2019efficientnet}, BigTransfer \cite{kolesnikov2020big}, and Vision Transformer \cite{dosovitskiy2020image}. These models required an input image size of 224x224 pixels. Moreover, the output embedding (feature) vectors were standardized to 256 dimensions per image across all models. Utilizing an assortment of pre-trained models, including both CNNs and transformers, the study sought to minimize the impact of any individual model's initial training on the overall outcomes.

\subsection{Base model configuration}
The embedding vectors (feature representations) extracted from the pre-trained models served as inputs for the base model. In this study, the base model is built using neural networks to perform binary classification. Considering the UQ technique employed in this study, a fully connected layers with an output layer equipped with a softmax function is employed. To determine the optimal architecture, the train set was first split into training and evaluation sets using a 70/30 ratio. The training set was used to train the models, while the evaluation set was reserved for evaluating their performance. Various architectures were explored by employing Keras Tuner's Hyperband algorithm. Keras Tuner, a library built on top of TensorFlow and Keras, automates the hyperparameter tuning process. Hyperband, an advanced tuning algorithm, dynamically allocates resources to train multiple models, stopping those that underperform early. The search space included variations in the number of hidden layers, ranging from one to four, and the number of neurons per layer, ranging from 16 to 512. Given the imbalanced nature of the dataset, class weights were computed to ensure balanced learning across classes. These weights were incorporated during the training process to prevent bias towards the majority class. The best-performing architecture was selected as the optimal base model for evaluating the proposed method. Additionally, MCD was employed during the prediction phase to estimate the model's uncertainty. In this study, the number of iterations for MCD was set to 100. 

In this study, three different pre-trained models were used, resulting in three sets of embedding vectors. For each set, a base model with its optimal architecture was determined. Table S1
in the Supplementary material summarizes the optimal architecture for each of the pre-trained models.

For a fair and unbiased comparison, both the traditional threshold-based method and the proposed U-SNN utilized the same trained base model.

\subsection{Meta-model configuration}
The meta-model is designed as a neural network, and its optimal architecture is determined using the Keras Tuner's Hyperband algorithm. This method, similar to that employed for the base model, efficiently explores the hyperparameter space to identify the most effective architecture. The search space for the meta-model includes variations in the number of hidden layers, ranging from one to four, and the number of neurons per layer, ranging from 16 to 512, allowing the model to adapt to the complexities of the data. Class weights are again utilized to address data imbalance, maintaining fairness in the learning process.

For the meta-model dataset generation, the confidence threshold is a crucial parameter that impacts the labeling of the data. In this study, five different thresholds were used: 0.05, 0.1, 0.2, and 0.4. These thresholds determine the cut-off points for labeling predictions as trustworthy or not, which in turn affects the training and performance of the meta-model. 

\section{Results and Discussion} \label{section:Discussion and Results}
Evaluating a model on a single test set might provide an initial performance snapshot, it does not guarantee reproducible results across different data splits. To address this variability and better gauge the model's generalization capabilities, the training and evaluation process is repeated 30 times. For each iteration, the dataset is randomly split into training and testing sets, with a test size randomly chosen between [20\%- 40\%] of the data. This repetitive approach ensures that the model is exposed to various data distributions, enable evaluation of the model's ability to generalize across various potential data distributions.

The classification performance of the base models, assessed independently of the uncertainty evaluation framework, is summarized in Table~\ref{tab:Base_model_performance_with_Entropy}. Three different pre-trained architectures—BiT, EfficientNetB0, and ViT—were evaluated based on their respective F1 scores and the AUC. Among the tested models, ViT yielded the highest classification performance, achieving an average F1 score of 86.95\% (±0.46) and an AUC of 94.21\% (±0.32). The BiT model followed closely, demonstrating an F1 score of 85.75\% (±0.40) and an AUC of 92.60\% (±0.33). EfficientNetB0 exhibited slightly lower performance, with an F1 score of 84.45\% (±0.46) and an AUC of 91.58\% (±0.23). These results confirmed that all base models provided strong predictive capabilities, suitable as foundational classifiers for subsequent uncertainty analysis. Notably, the consistently high AUC values across all three models indicate robust discrimination ability, further supporting their effectiveness in the underlying classification task.

\begin{table}[]
\caption{Summary of base models' performance across various pre-trained models}
\label{tab:Base_model_performance_with_Entropy}
\resizebox{\columnwidth}{!}{%
\begin{tabular}{llcc}
\hline
\multicolumn{1}{c}{Pre-trained Model} & \multicolumn{1}{c}{UQ} & F1 score           & AUC                \\ \hline
BiT                                   & MCD                    & \%85.7497 ± 0.3985 & \%92.5964 ± 0.3318 \\
EfficientNetB0                        & MCD                    & \%84.4497 ± 0.4569 & \%91.5801 ± 0.2291 \\
ViT                                   & MCD                    & \%86.9507 ± 0.4568 & \%94.2096 ± 0.3157 \\ \hline
\end{tabular}%
}
\end{table}

As the proposed U-SNN approach incorporates an additional meta-model specifically designed to assess the trustworthiness of predictions, the overall reliability of the U-SNN directly depends on the performance of this meta-model. Accordingly, before comparing the uncertainty performance metrics between the U-SNN and the traditional threshold-based method, the classification performance of the meta-model was evaluated separately. This evaluation aimed to verify the meta-model's capability to correctly distinguish trustworthy predictions (correctly identifying correct and confident predictions) from untrustworthy predictions (either incorrect or uncertain). It should be noted that these results pertain exclusively to the proposed U-SNN, as the traditional threshold-based method does not include a meta-model. The performance of the meta-model is influenced by the confidence threshold, as determined by the target variable definition in Equation~\ref{Eq:Meta_train_set_label}. Table \ref{tab:Bmeta_model_performance_with_Entropy_th01} illustrates the meta-model performance across three as pre-trained models at confidence threshold 0.1. Supplementary Table S.2 presents a summary of the meta-model's error-based performance results across confidence thresholds of 0.05, 0.2, and 0.4.

Across the three pre-trained architectures examined—BiT, EfficientNetB0, and ViT—the meta-model consistently demonstrated robust performance, with high F1 scores and exceptionally high AUC values. Specifically, the ViT-based meta-model yielded the highest performance, achieving an F1 score of 97.48\% (±0.32) and an AUC of 99.63\% (±0.11). The BiT-based meta-model closely followed, recording an F1 score of 96.65\% (±0.35) and an AUC of 99.50\% (±0.09), while the EfficientNetB0-based meta-model displayed slightly lower but still strong results, with an F1 score of 96.37\% (±0.45) and an AUC of 99.47\% (±0.13). These results indicate that the meta-model reliably distinguishes between trustworthy and untrustworthy predictions.

Following, a direct comparison of uncertainty metrics between the U-SNN framework and the traditional threshold-based approach is presented in Table \ref{tab:comparison_USNN_Threshold_based}.

\begin{table}[]
\caption{Summary of meta-models' performance across various pre-trained models at the confidence threshold 0.1}
\label{tab:Bmeta_model_performance_with_Entropy_th01}
\resizebox{\columnwidth}{!}{%
\begin{tabular}{llll}
\hline
\multicolumn{1}{c}{Model} & \multicolumn{1}{c}{UQ} & \multicolumn{1}{c}{F1 score} & \multicolumn{1}{c}{AUC} \\ \hline
BiT                       & MCD                    & \%96.6539 ± 0.3515           & \%99.5043 ± 0.0949      \\
EfficientNetB0            & MCD                    & \%96.3704 ± 0.4546           & \%99.4674 ± 0.1265      \\
ViT                       & MCD                    & \%97.4823 ± 0.3232           & \%99.6293 ± 0.1102      \\ \hline
\end{tabular}%
}
\end{table}

To start, let's first examine the results at confidence threshold 0.1. Figure \ref{fig:USNN_performance_comparison_th01} illustrates a comparative analysis of the U-SNN and the threshold-based method at a confidence threshold of 0.1. At a confidence threshold of 0.1, both the U-SNN and threshold-based approaches displayed similar CR across all models. This observation suggests that both methods were equally assertive in labeling predictions as confident. However, similarity in CR values does not inherently indicate reliability, as it does not account for the correctness of these confident predictions. The distinction between the two methods becomes evident when examining the FCR and CE. FCR measures how often is the model confidently incorrect among all predictions. The U-SNN consistently demonstrated a substantial reduction in FCR compared to the threshold-based approach. For the BiT model, U-SNN achieving a 55.4\% reduction in FCR and marking a 48.4\% and 47.7\% improvement in FCR for EfficientNetB0 and ViT, respectively. This indicates that U-SNN was considerably more effective in preventing confidently incorrect predictions. The implication is significant: in high-stakes scenarios, such as medical diagnosis, confidently incorrect predictions can mislead operators into trusting erroneous results. By substantially reducing FCR, the U-SNN framework mitigates this critical risk.

Similarly, the CE metric further highlights the robustness of U-SNN by assessing the proportion of errors among confident predictions. U-SNN achieved a 55.9\% reduction in CE for the BiT model, along with reductions of 50.9\% and 48.0\% for EfficientNetB0 and ViT, respectively. These improvements indicate that when U-SNN flags a prediction as confident, there is a significantly higher likelihood that it is indeed correct. In contrast, the higher CE in the threshold-based approach reflects a vulnerability, where confidently incorrect predictions are more prevalent, potentially eroding trust in the system.

The UR remained relatively similar between the methods, suggesting that both approaches referred a similar proportion of predictions for further review. However, the key distinction arises in the RR metric, which captures the inefficiency of unnecessary reviews: For BiT, RR dropped from 97.40\% to 85.92\%, a 11.8\% reduction in unnecessary referrals. For EfficientNetB0 reflecting a 12.3\% improvement and For ViT  representing a 13.9\% reduction in RR. Although UR rates were similar, the substantial reduction in RR demonstrates that U-SNN was far more selective and efficient in its referrals. This has significant practical implications: unnecessary referrals increase workload, delay decision-making, and strain resources. By reducing RR, the U-SNN approach enhances operational efficiency, ensuring that human reviewers focus only on genuinely uncertain or complex cases that require further analysis.

\begin{table*}
\centering
\caption{Comparison of traditional threshold-base model and proposed method (U-SNN)}
\label{tab:comparison_USNN_Threshold_based}
\resizebox{\linewidth}{!}{%
\begin{tblr}{
  row{1} = {c},
  cell{2}{1} = {r=2}{},
  cell{2}{2} = {r=6}{c},
  cell{4}{1} = {r=2}{},
  cell{6}{1} = {r=2}{},
  cell{8}{1} = {r=2}{},
  cell{8}{2} = {r=6}{c},
  cell{10}{1} = {r=2}{},
  cell{12}{1} = {r=2}{},
  cell{14}{1} = {r=2}{},
  cell{14}{2} = {r=6}{c},
  cell{16}{1} = {r=2}{},
  cell{18}{1} = {r=2}{},
  cell{20}{1} = {r=2}{},
  cell{20}{2} = {r=6}{c},
  cell{22}{1} = {r=2}{},
  cell{24}{1} = {r=2}{},
  hline{1-2,26} = {-}{},
  hline{4,6,10,12,16,18,22,24} = {1,3-8}{dashed,Gray},
  hline{8,14,20} = {-}{Gray},
}
Pretrained & Confidence Threshold & Approach & CR & FCR & CE & UR & RR\\
BiT & 0.05 & U-SNN & 29.4514 ± 4.4865 & 0.2081 ± 0.0749 & 0.6902 ± 0.1652 & 70.5486 ± 4.4865 & 87.4771 ± 0.6375\\
 &  & Threshold-Based & 29.4052 ± 4.6884 & 0.7444 ± 0.2518 & 2.5239 ± 0.7174 & 70.5948 ± 4.6884 & 97.4181 ± 0.6319\\
EfficientNetB0 &  & U-SNN & 27.4149 ± 3.7112 & 0.2182 ± 0.0897 & 0.7724 ± 0.2322 & 72.5851 ± 3.7112 & 86.7379 ± 0.5236\\
 &  & Threshold-Based & 26.9223 ± 4.3993 & 0.7121 ± 0.2046 & 2.6735 ± 0.7823 & 73.0777 ± 4.3993 & 97.2557 ± 0.5887\\
ViT &  & U-SNN & 39.3139 ± 5.8126 & 0.2212 ± 0.0981 & 0.5408 ± 0.1706 & 60.6861 ± 5.8126 & 86.4584 ± 1.0562\\
 &  & Threshold-Based & 39.8193 ± 5.1586 & 0.7203 ± 0.2274 & 1.7866 ± 0.4357 & 60.1807 ± 5.1586 & 98.1765 ± 0.4355\\
BiT & 0.1 & U-SNN & 38.7992 ± 5.0042 & 0.4395 ± 0.1458 & 1.107 ± 0.2486 & 61.2008 ± 5.0042 & 85.9213 ± 0.828\\
 &  & Threshold-Based & 39.1187 ± 4.9228 & 0.9857 ± 0.2939 & 2.5228 ± 0.6806 & 60.8813 ± 4.9228 & 97.4005 ± 0.6731\\
EfficientNetB0 &  & U-SNN & 36.1502 ± 3.9059 & 0.4755 ± 0.1488 & 1.2907 ± 0.2857 & 63.8498 ± 3.9059 & 85.3193 ± 0.6299\\
 &  & Threshold-Based & 35.7458 ± 4.5747 & 0.9323 ± 0.2528 & 2.6277 ± 0.7215 & 64.2542 ± 4.5747 & 97.2246 ± 0.5881\\
ViT &  & U-SNN & 48.3345 ± 5.5188 & 0.4623 ± 0.1625 & 0.9315 ± 0.2372 & 51.6655 ± 5.5188 & 84.5481 ± 1.2343\\
 &  & Threshold-Based & 48.841 ± 4.9201 & 0.8828 ± 0.2492 & 1.7931 ± 0.413 & 51.159 ± 4.9201 & 98.1729 ± 0.459\\
BiT & 0.2 & U-SNN & 52.9509 ± 5.2666 & 1.1213 ± 0.2945 & 2.0862 ± 0.3533 & 47.0491 ± 5.2666 & 83.098 ± 1.1284\\
 &  & Threshold-Based & 53.8595 ± 4.8049 & 1.3808 ± 0.3561 & 2.5623 ± 0.608 & 46.1405 ± 4.8049 & 97.4307 ± 0.7138\\
EfficientNetB0 &  & U-SNN & 50.0759 ± 4.1463 & 1.2038 ± 0.2583 & 2.3804 ± 0.339 & 49.9241 ± 4.1463 & 82.6654 ± 0.8199\\
 &  & Threshold-Based & 50.18 ± 4.3741 & 1.3476 ± 0.3392 & 2.7034 ± 0.7264 & 49.82 ± 4.3741 & 97.2567 ± 0.5432\\
ViT &  & U-SNN & 60.0983 ± 4.7476 & 1.0308 ± 0.2663 & 1.692 ± 0.3233 & 39.9017 ± 4.7476 & 81.4207 ± 1.411\\
 &  & Threshold-Based & 60.679 ± 4.5586 & 1.1108 ± 0.2762 & 1.8227 ± 0.3947 & 39.321 ± 4.5586 & 98.205 ± 0.5112\\
BiT & 0.4 & U-SNN & 75.6858 ± 3.5126 & 3.3056 ± 0.4536 & 4.3513 ± 0.4165 & 24.3142 ± 3.5126 & 76.3087 ± 1.4808\\
 &  & Threshold-Based & 76.7593 ± 3.5632 & 1.9887 ± 0.4471 & 2.5922 ± 0.5726 & 23.2407 ± 3.5632 & 97.5396 ± 0.7978\\
EfficientNetB0 &  & U-SNN & 72.9876 ± 3.8683 & 3.4535 ± 0.4571 & 4.7129 ± 0.3947 & 27.0124 ± 3.8683 & 76.2099 ± 1.5066\\
 &  & Threshold-Based & 74.1529 ± 3.4815 & 1.9709 ± 0.4456 & 2.6669 ± 0.6355 & 25.8471 ± 3.4815 & 97.1108 ± 0.5627\\
ViT &  & U-SNN & 77.2759 ± 3.3395 & 2.669 ± 0.4377 & 3.4373 ± 0.4289 & 22.7241 ± 3.3395 & 74.6181 ± 1.7885\\
 &  & Threshold-Based & 78.0048 ± 3.1074 & 1.4212 ± 0.3269 & 1.8194 ± 0.3991 & 21.9952 ± 3.1074 & 98.2178 ± 0.5568
\end{tblr}
}
\end{table*}

\begin{figure}[]
  \centering
  \captionsetup[subfloat]{font=tiny}
  \subfloat[][CR]{\includegraphics[width=0.5\columnwidth]{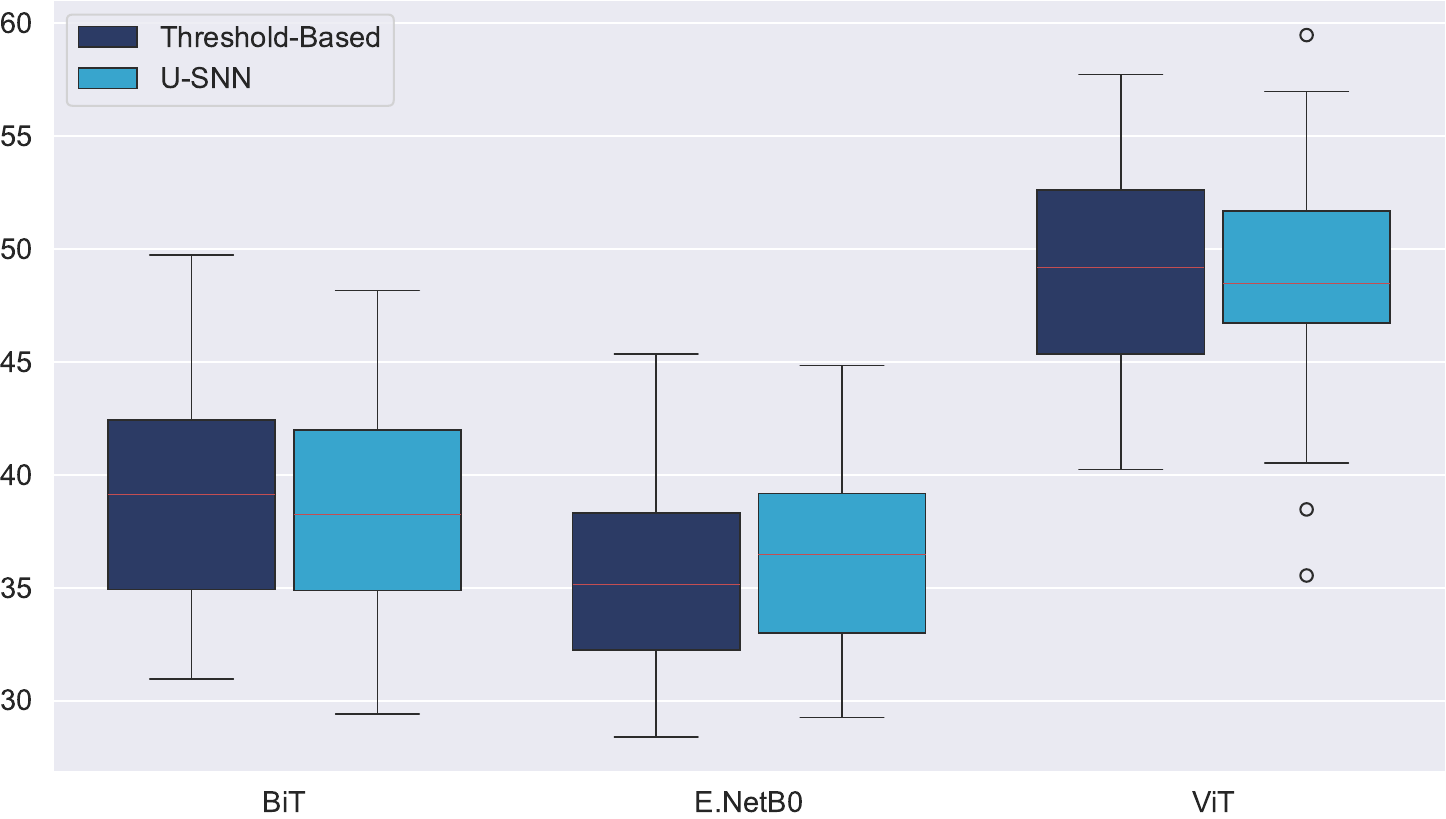}}
  \subfloat[][FCR]{\includegraphics[width=0.5\columnwidth]{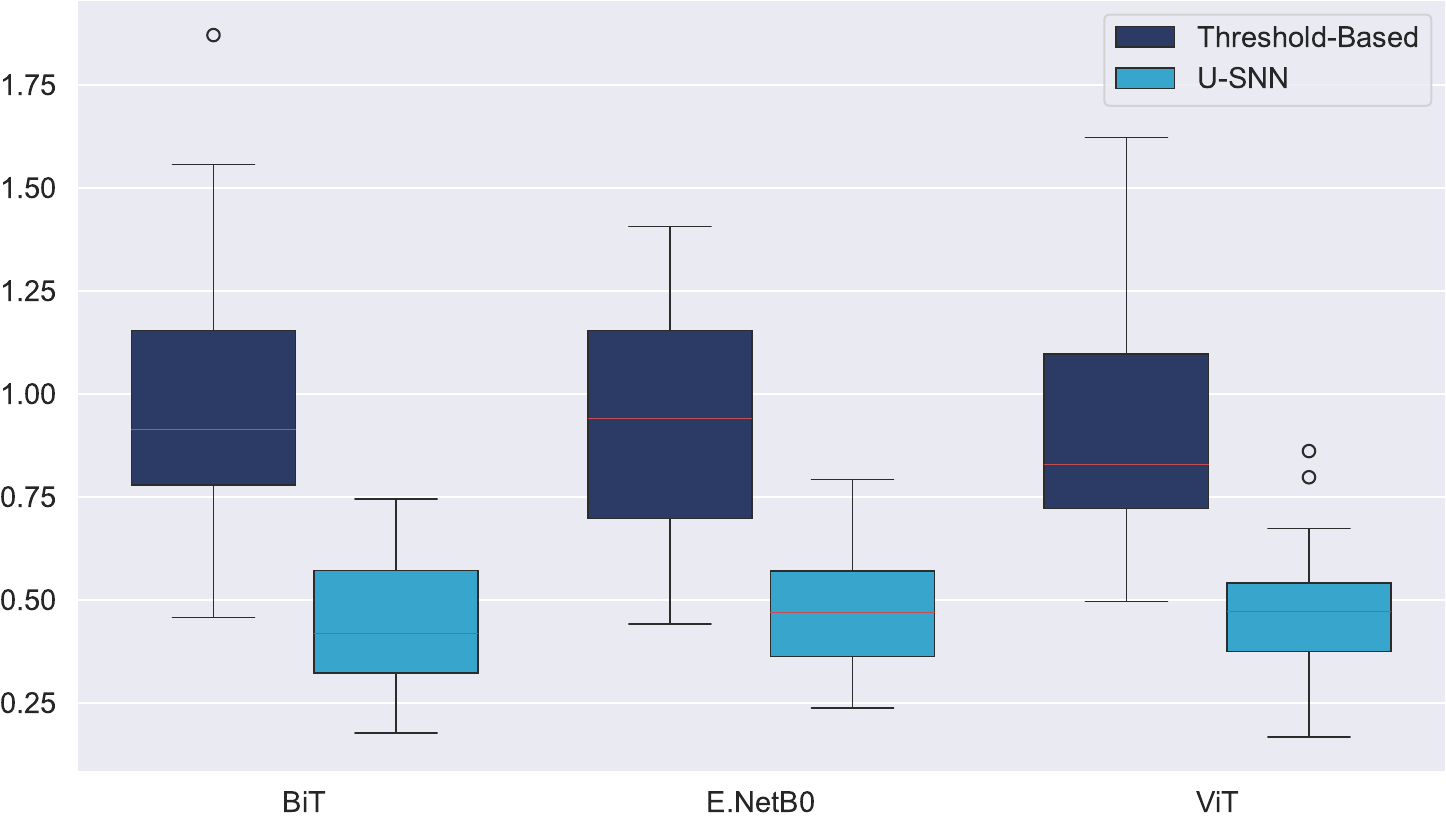}}\par
  \subfloat[][CE]{\includegraphics[width=0.5\columnwidth]{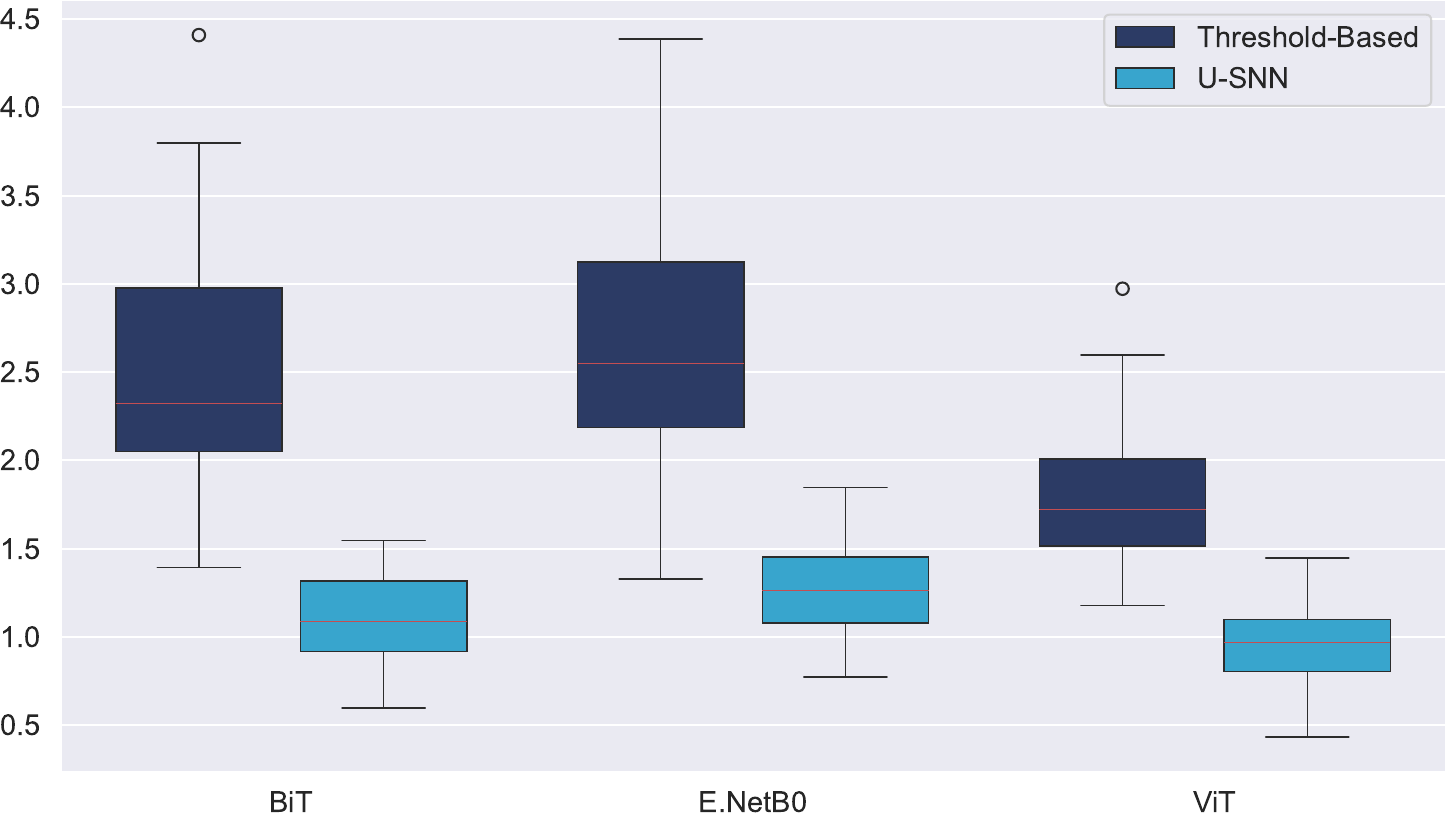}}
  \subfloat[][UR]{\includegraphics[width=0.5\columnwidth]{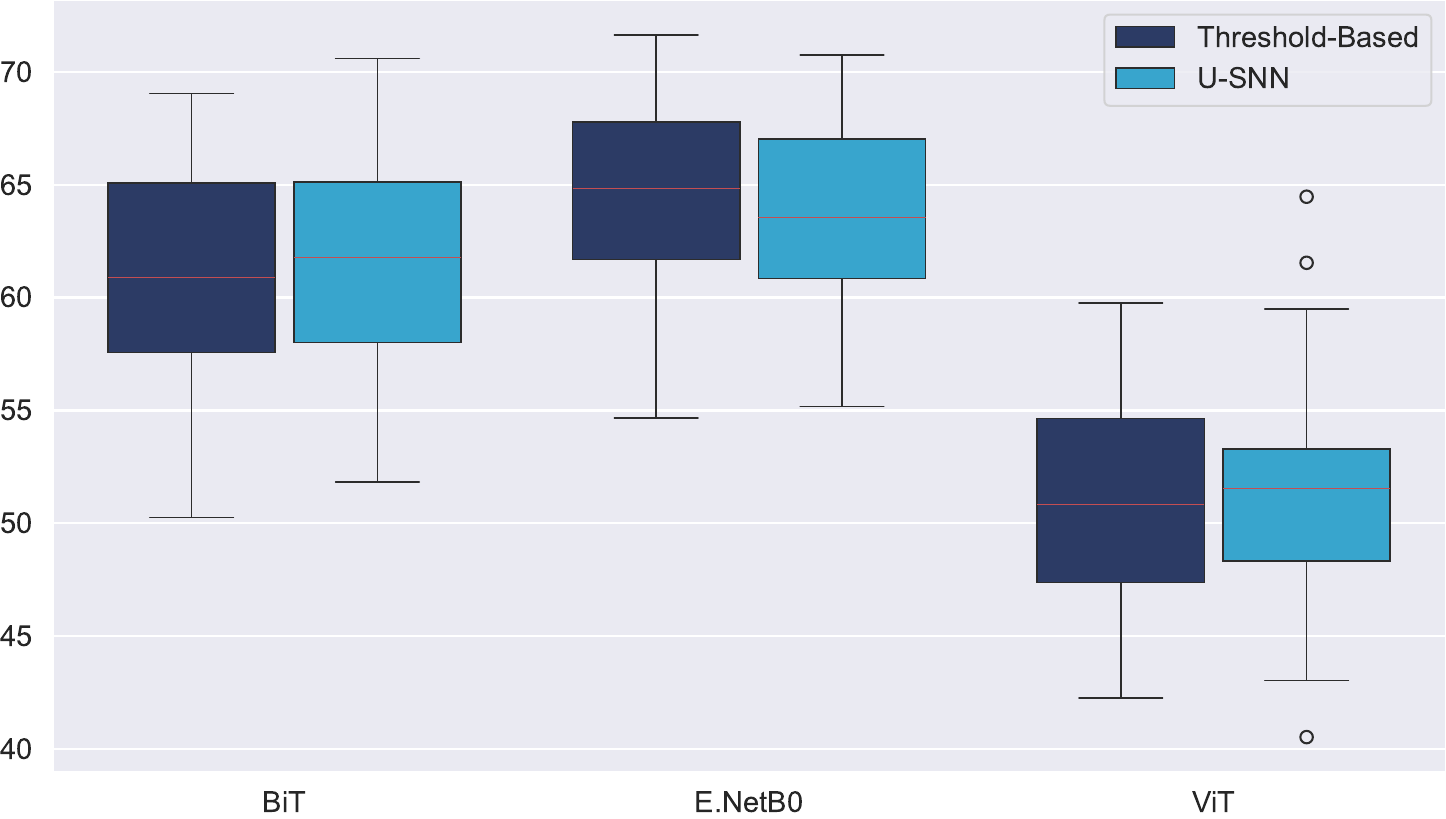}}\par
  \subfloat[][RR]{\includegraphics[width=0.5\columnwidth]{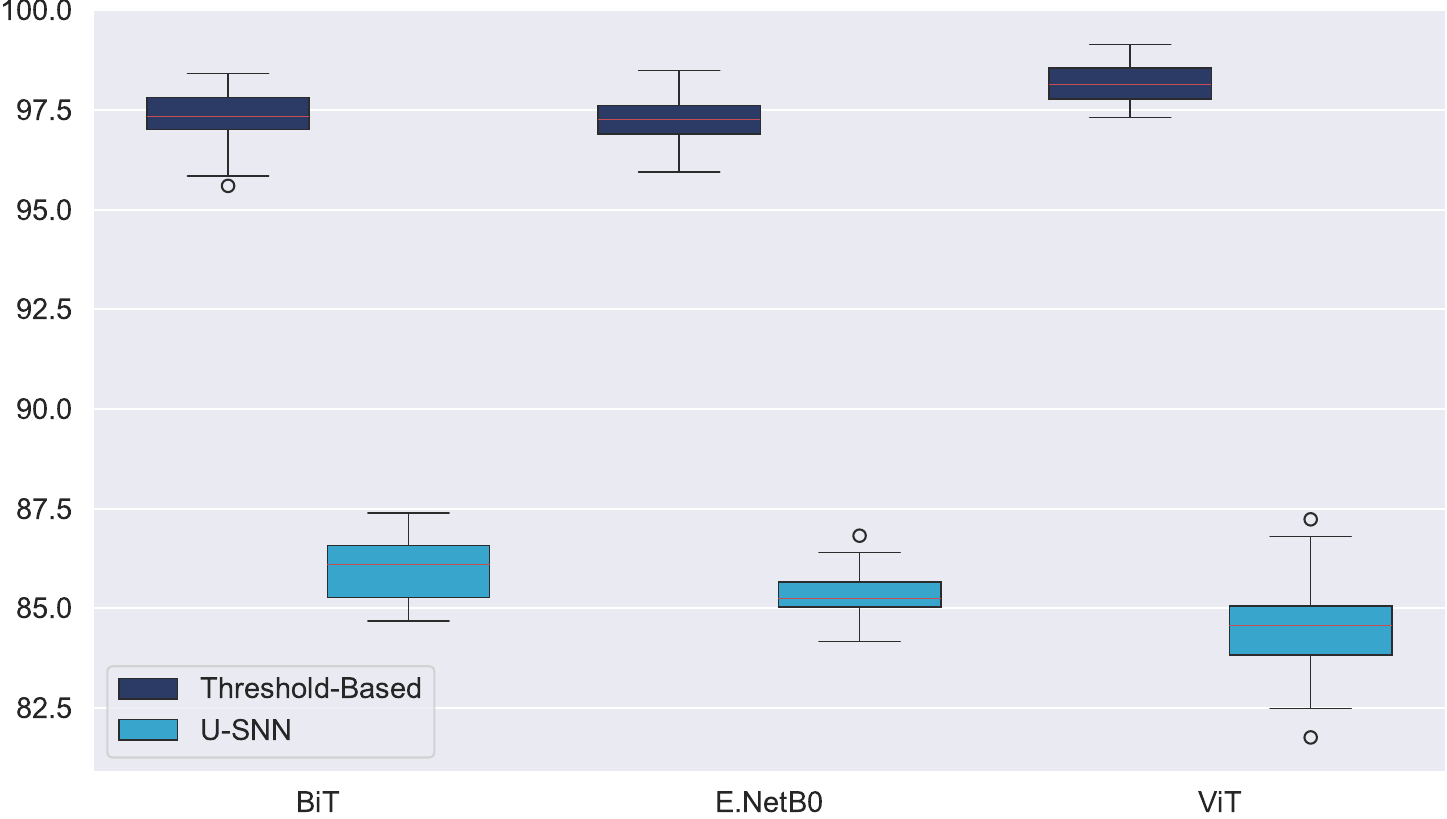}}
  \caption{Comparison of Uncertainty-Informed Criteria Across Pre-trained Models at a Confidence Threshold of 0.1, U-SNN and Tradiditonal Threshold-based method.}
  \label{fig:UI_Citeria_Comparison_Inputs}
\end{figure}

\begin{figure}[]
  \centering
  \captionsetup[subfloat]{font=tiny}
  \subfloat[][CR]{\includegraphics[width=0.5\columnwidth]{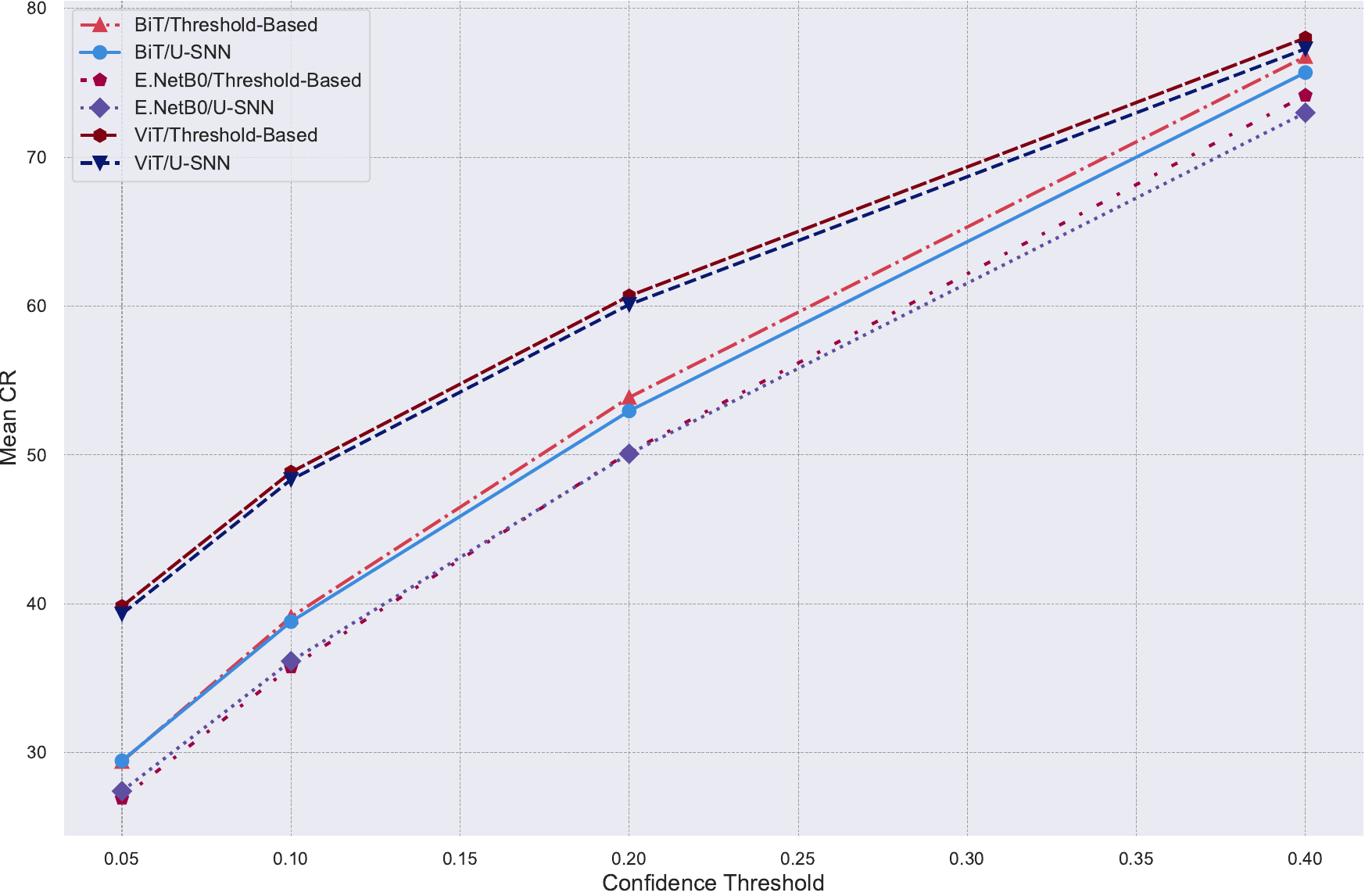}}
  \subfloat[][FCR]{\includegraphics[width=0.5\columnwidth]{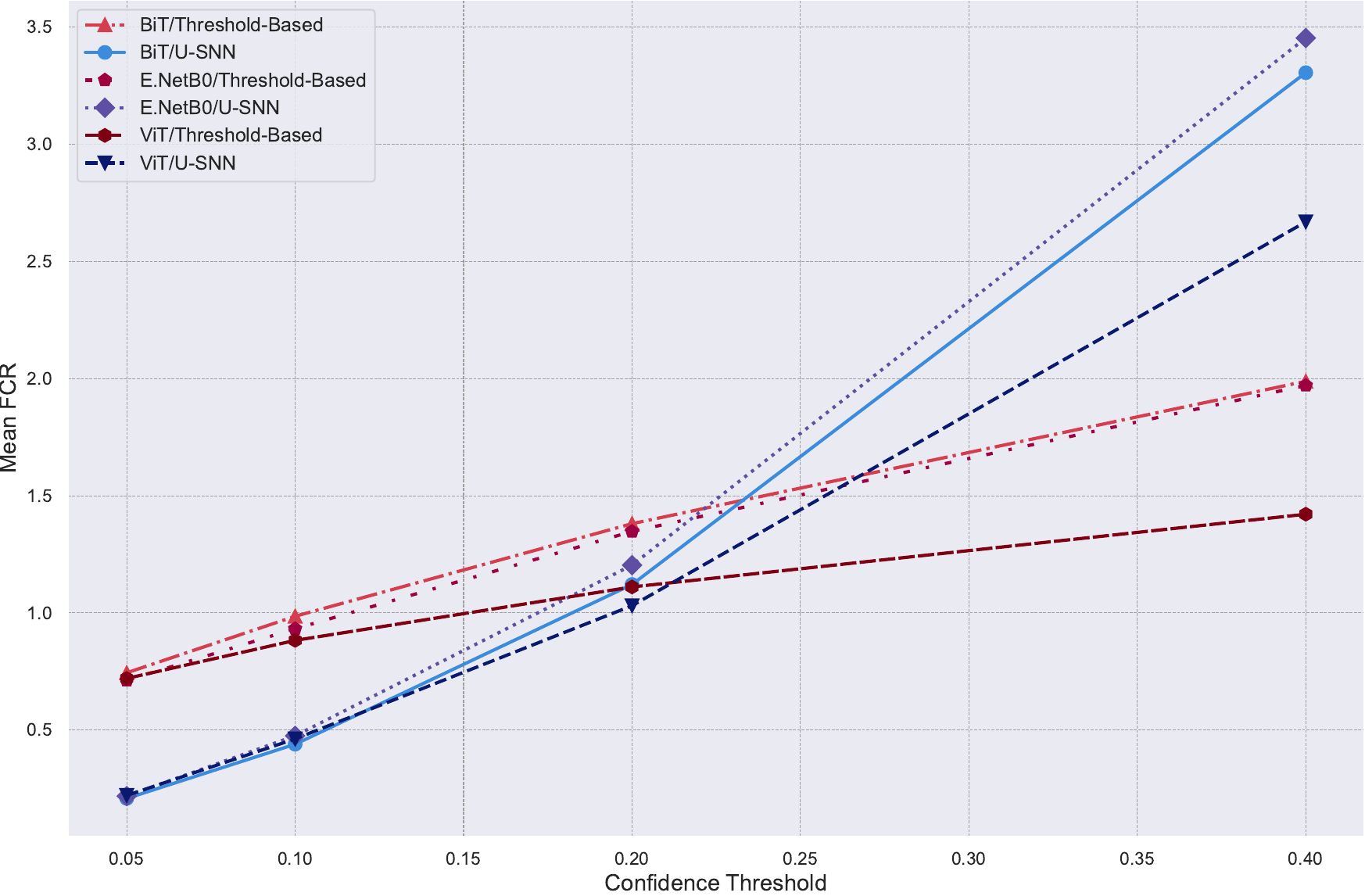}}\par
  \subfloat[][CE]{\includegraphics[width=0.5\columnwidth]{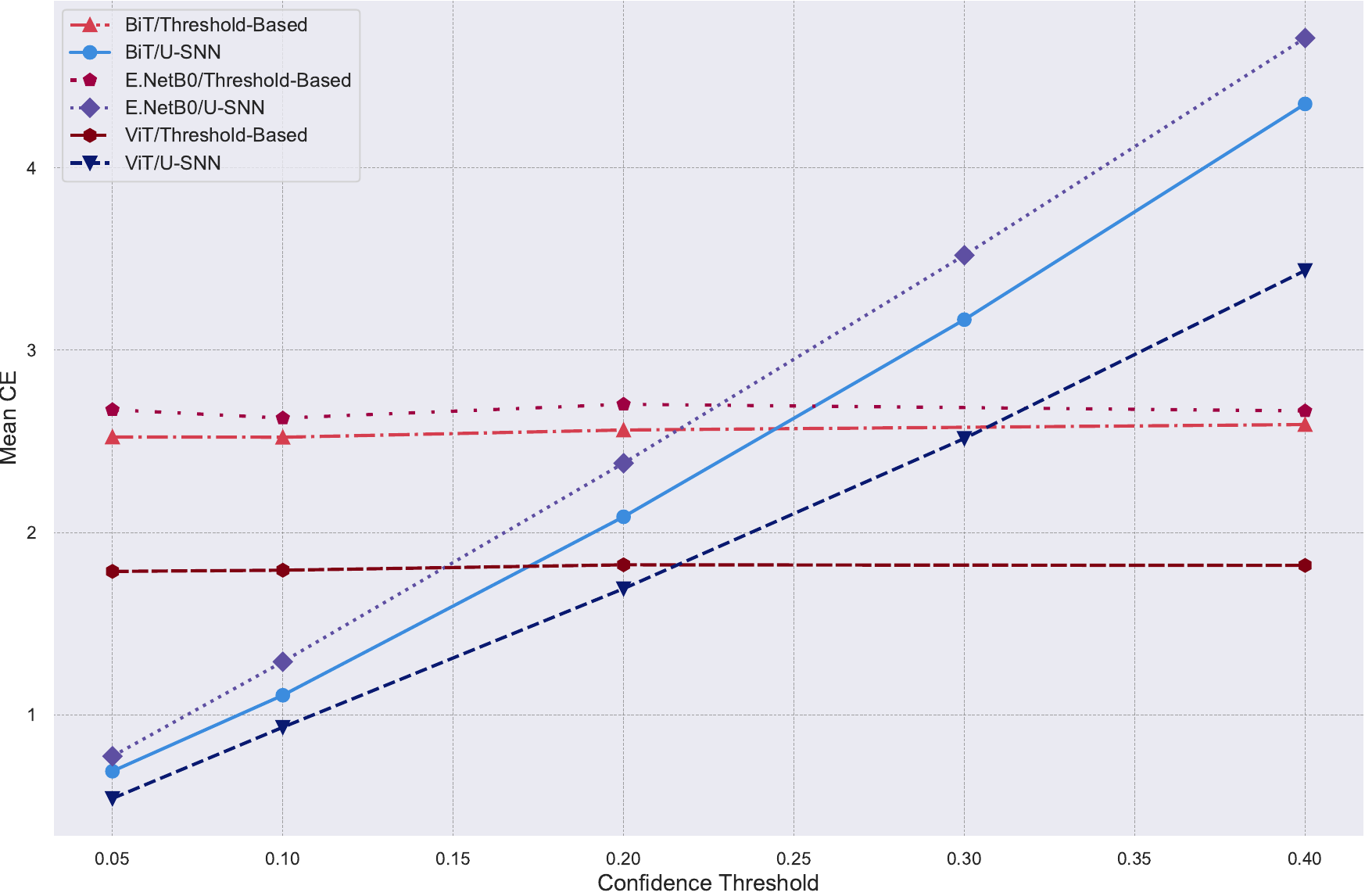}}
  \subfloat[][UR]{\includegraphics[width=0.5\columnwidth]{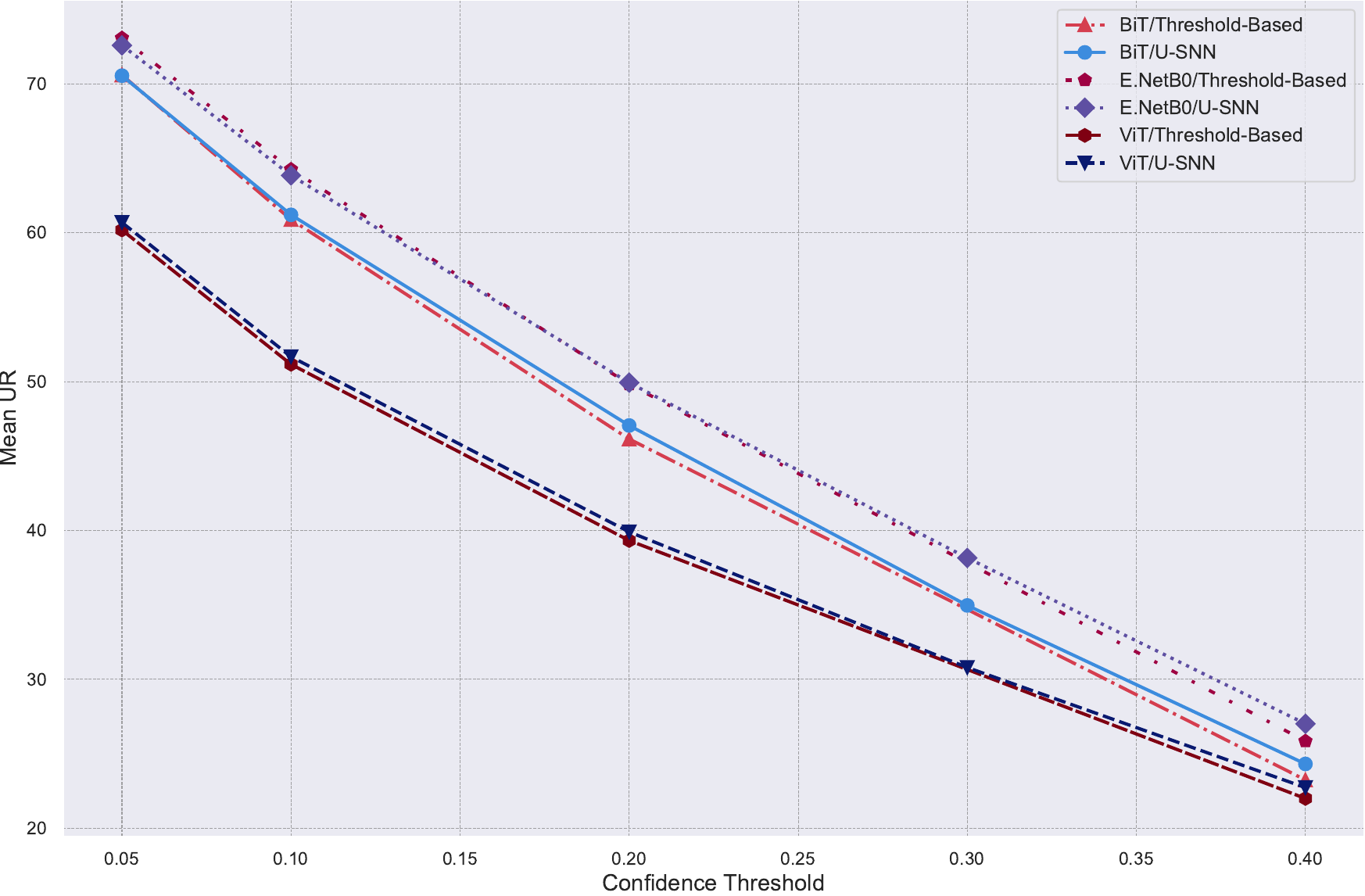}}\par
  \subfloat[][RR]{\includegraphics[width=0.5\columnwidth]{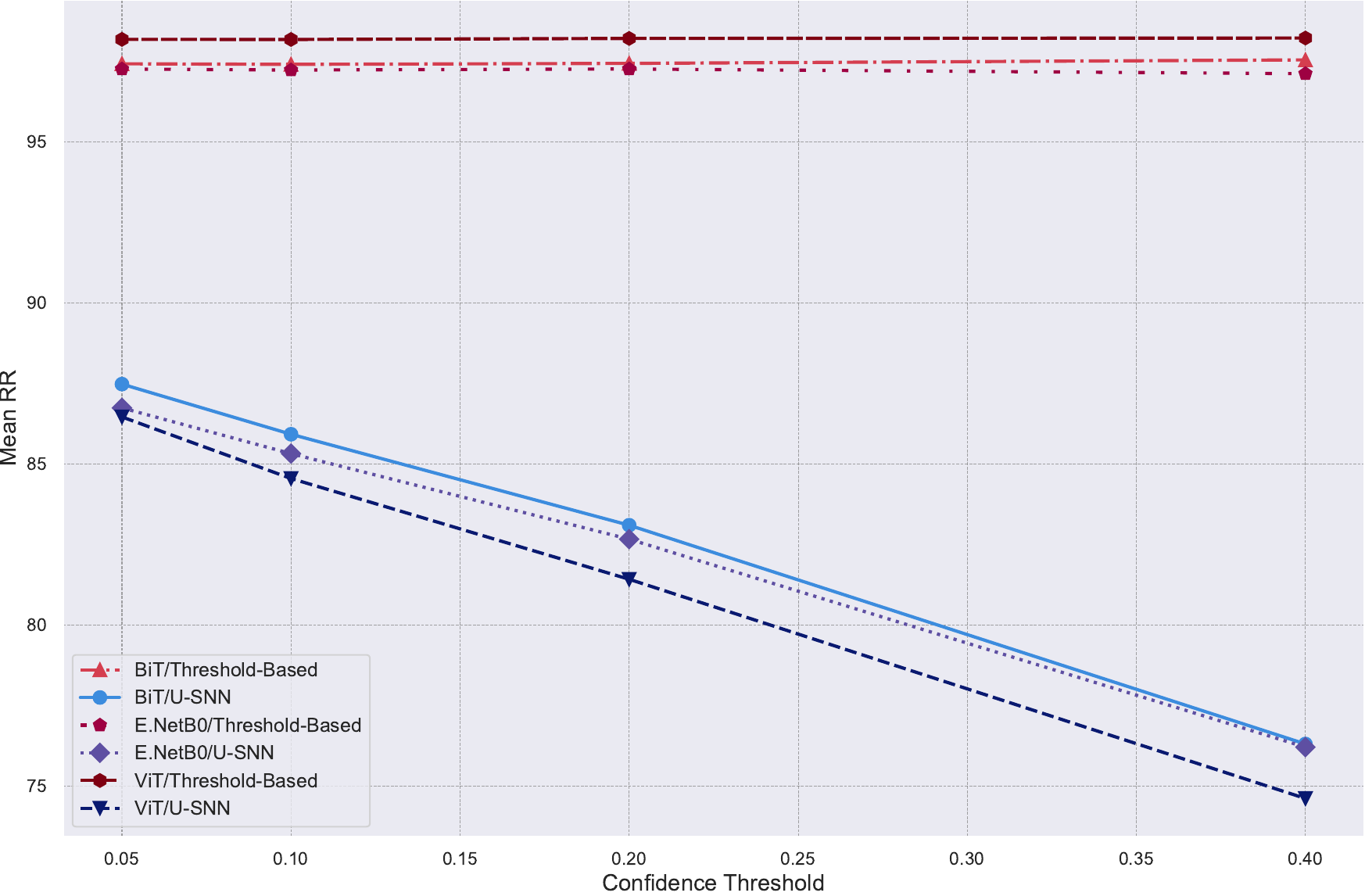}}
  \caption{Trend Analysis of Uncertainty-Informed Criteria Across Various Pre-trained Models at Incremental Confidence Thresholds from 0.05 to 0.4.}
  \label{fig:UI_Citeria_Trend_of_various_thresholds}
\end{figure}

To evaluate the comparative performance of the U-SNN and traditional threshold-based approaches, trends in key metrics were analyzed as the confidence threshold increased from 0.05 to 0.4. Figure \ref{fig:UI_Citeria_Trend_of_various_thresholds} illustrates the trends in evaluation metrics for both the proposed U-SNN and the traditional threshold-based approaches across three pre-trained models (BiT, EfficientNetB0, and ViT), with confidence thresholds varying from 0.05 to 0.4. Both methods exhibited a consistent increase in CR as the threshold increased. This is expected, as higher thresholds result in a greater number of predictions being classified as confident. While CR trends were similar, subsequent metrics revealed significant differences in prediction correctness and referral efficiency.

At lower thresholds (0.05 to 0.2), U-SNN consistently outperformed the threshold-based method in minimizing confidently incorrect predictions. For example, BiT’s FCR in U-SNN rose from 0.21\% at 0.05 to 1.12\% at 0.2, whereas the threshold-based approach increased from 0.74\% to 1.38\% over the same range. However, at the highest threshold of 0.4, this pattern reversed. The threshold-based approach achieved a lower FCR of 1.99\%, compared to U-SNN's 3.31\%. Similar reversals were observed in EfficientNetB0 and ViT.

A similar trend was observed in CE, which also favored U-SNN at lower thresholds but reversed at higher thresholds. For BiT, U-SNN increased CE from 0.69\% at 
0.05 to 2.09\% at 0.2, consistently maintaining lower values than the threshold-based method. However, at 0.4, the traditional approach yielded a lower CE of 2.59\%, compared to U-SNN’s 4.35\%.

FCR and CE results suggest that as the threshold becomes more lenient, the U-SNN’s  performance degrades. This may be due to the meta-model's diminishing ability to differentiate trustworthy predictions when a large proportion of predictions are automatically deemed confident. Therefore, the U-SNN framework is particularly advantageous in moderate-threshold scenarios, where balancing certainty with correctness is critical. However, when the system operates under very high thresholds, where almost all predictions are labeled as confident, the simpler threshold-based approach may provide tighter control over confidently incorrect predictions.

Although the threshold-based approach exhibited lower FCR and CE at the highest evaluated threshold (0.4), this outcome is less critical when considering practical applications. In uncertainty quantification, thresholds above 0.5 are generally discouraged, as they increase the risk of accepting low-confidence predictions as confident, particularly in high-stakes contexts such as medical diagnosis. Lower thresholds (e.g., 0.05 to 0.2) are scientifically favored as they encourage more conservative and reliable uncertainty estimation. Importantly, U-SNN demonstrated superior performance across these practically relevant lower thresholds, substantially reducing confidently incorrect predictions and redundant referrals. Thus, while U-SNN's performance slightly degrades at higher thresholds, this scenario is less operationally relevant, and the framework remains superior in the thresholds that matter most for trustworthy decision-making.

In terms of UR and RR, both methods exhibited a decreasing UR trend with rising thresholds, indicating that fewer predictions were flagged for review. This consistent trend suggests that both methods became less conservative in uncertainty classification as the threshold increased. The most notable advantage of U-SNN was observed in the RR metric. Across all thresholds, U-SNN consistently achieved substantially lower RR values. This highlights that U-SNN was consistently more efficient in minimizing unnecessary referrals, ensuring that human reviewers were focused on genuinely uncertain cases. The threshold-based method, however, remained inefficient, flagging a high number of already-correct predictions for unnecessary review.

\subsection{Ablation Study}

In this ablation study, the impact of incorporating PE as an input to the meta-model was examined. The meta-model was trained under two conditions: one in which PE was included alongside the feature representations from the pre-trained model, and another in which only the feature representations were used as input. The goal is to determine whether the inclusion of PE contributes to more reliable and efficient uncertainty-aware decision-making within the U-SNN framework.

Table \ref{tab:meta_model_performance_Excl_PE_Th01} presents the error-based performance results of the meta-model excluding PE at confidence threshold 0.1. Supplementary Table S.2 provides a summary of the error-based performance results, comparing models with and without PE across various confidence thresholds values.

% , while Supplementary Figures S.1 to S.3 display the performance comparisons at confidence threshold values of 0.05, 0.2, and 0.4, respectively.

The comparison between meta-models that include PE as input and those that exclude it reveals a consistent trend across both F1 scores and AUC values. Meta-models that incorporate PE as part of their input tend to achieve higher F1 scores, as well as AUC, across all pre-trained models. The underlying reason for these improvements is that PE provides the meta-model with crucial contextual information about the certainty of predictions. Without PE, the meta-model relies solely on feature representations, which may not fully capture the uncertainty associated with each prediction. By incorporating PE, the model can adjust its predictions based on its confidence, leading to more accurate and reliable performance.

\begin{table}[]
\centering
\caption{Performance Summary of Meta-Models Across Pre-trained Models, Excluding PE as Input at Confidence Threshold 0.1}
\label{tab:meta_model_performance_Excl_PE_Th01}
\resizebox{\columnwidth}{!}{%
\begin{tabular}{llll}
\hline
\multicolumn{1}{c}{Model} & \multicolumn{1}{c}{UQ} & \multicolumn{1}{c}{F1 score} & \multicolumn{1}{c}{AUC} \\ \hline
BiT & MCD & \%91.9709 ± 1.0608 & \%97.9825 ± 0.5543 \\
EfficientNetB0 & MCD & \%91.1689 ± 1.1750 & \%97.6144 ± 0.7021 \\
ViT & MCD & \%91.0680 ± 1.1215 & \%97.4382 ± 0.5998 \\ \hline
\end{tabular}%
}
\end{table}

Improvement in the meta-model's performance also enhances trust-informed decision-making. Table \ref{tab:certainty_result_Exclud_PE} presents the results of the proposed  metrics excluding PE.

\begin{figure}[]
  \centering
  \captionsetup[subfloat]{font=tiny}
  \subfloat[][CR]{\includegraphics[width=0.5\columnwidth]{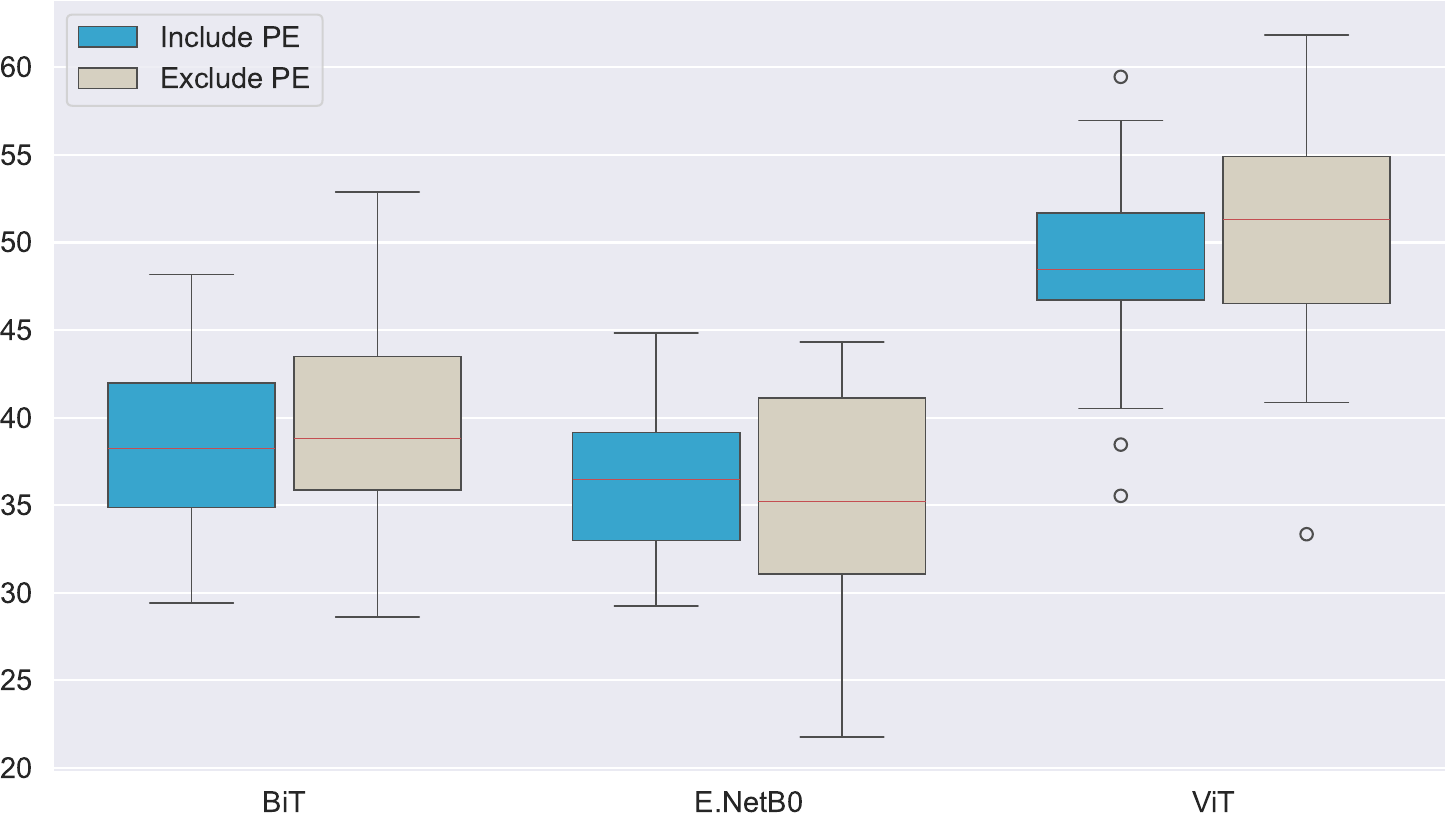}}
  \subfloat[][FCR]{\includegraphics[width=0.5\columnwidth]{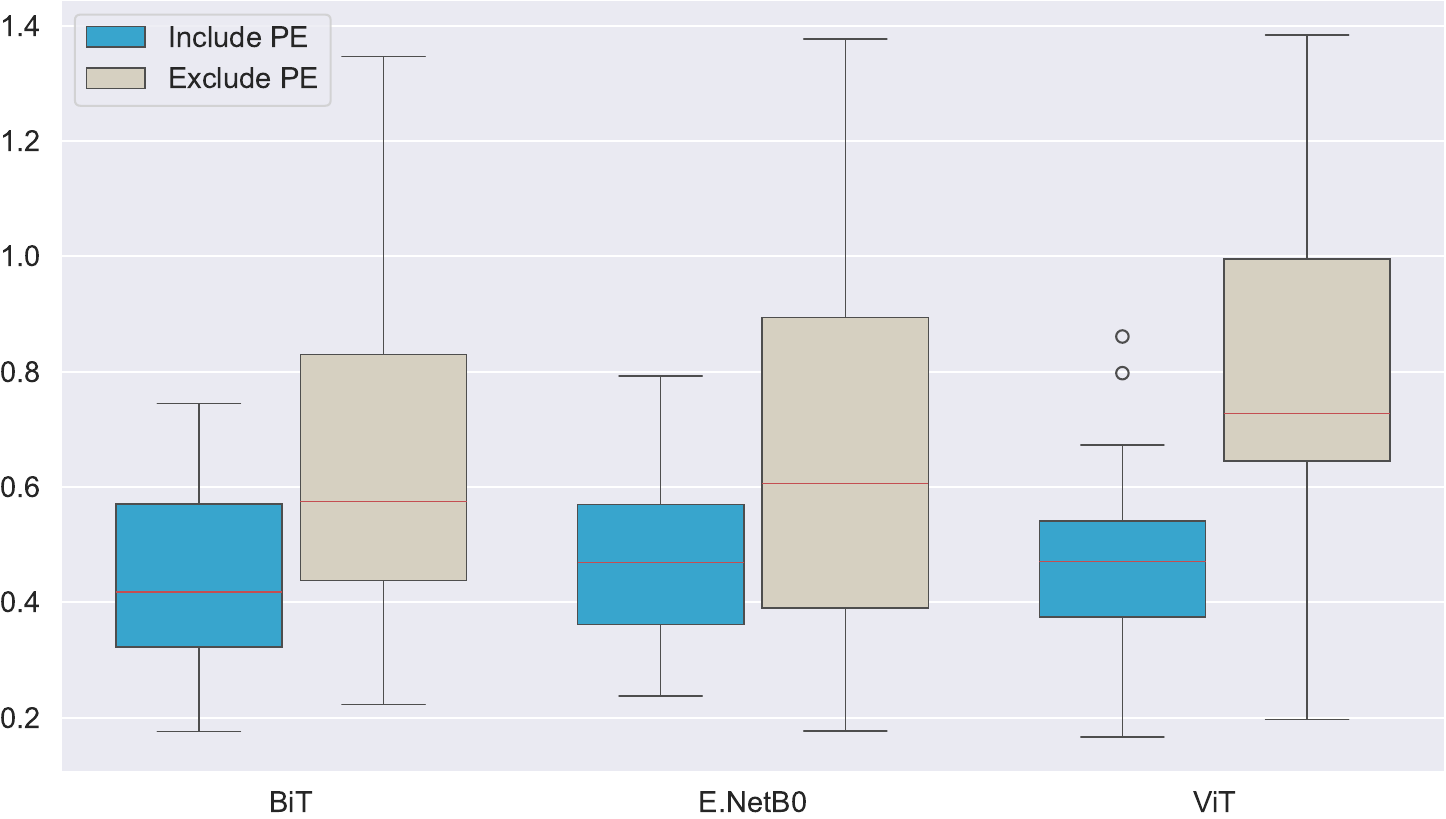}}\par
  \subfloat[][CE]{\includegraphics[width=0.5\columnwidth]{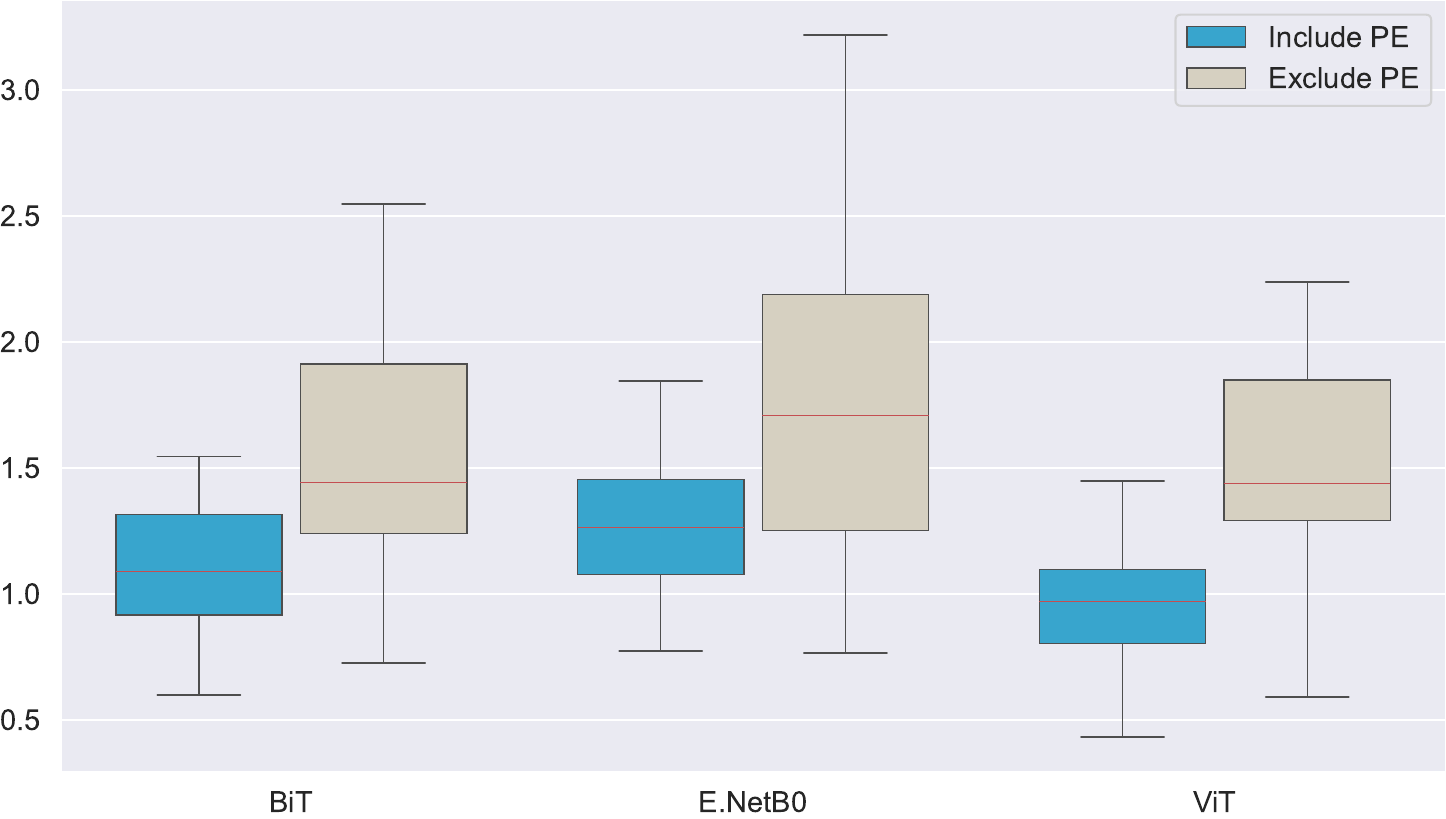}}
  \subfloat[][UR]{\includegraphics[width=0.5\columnwidth]{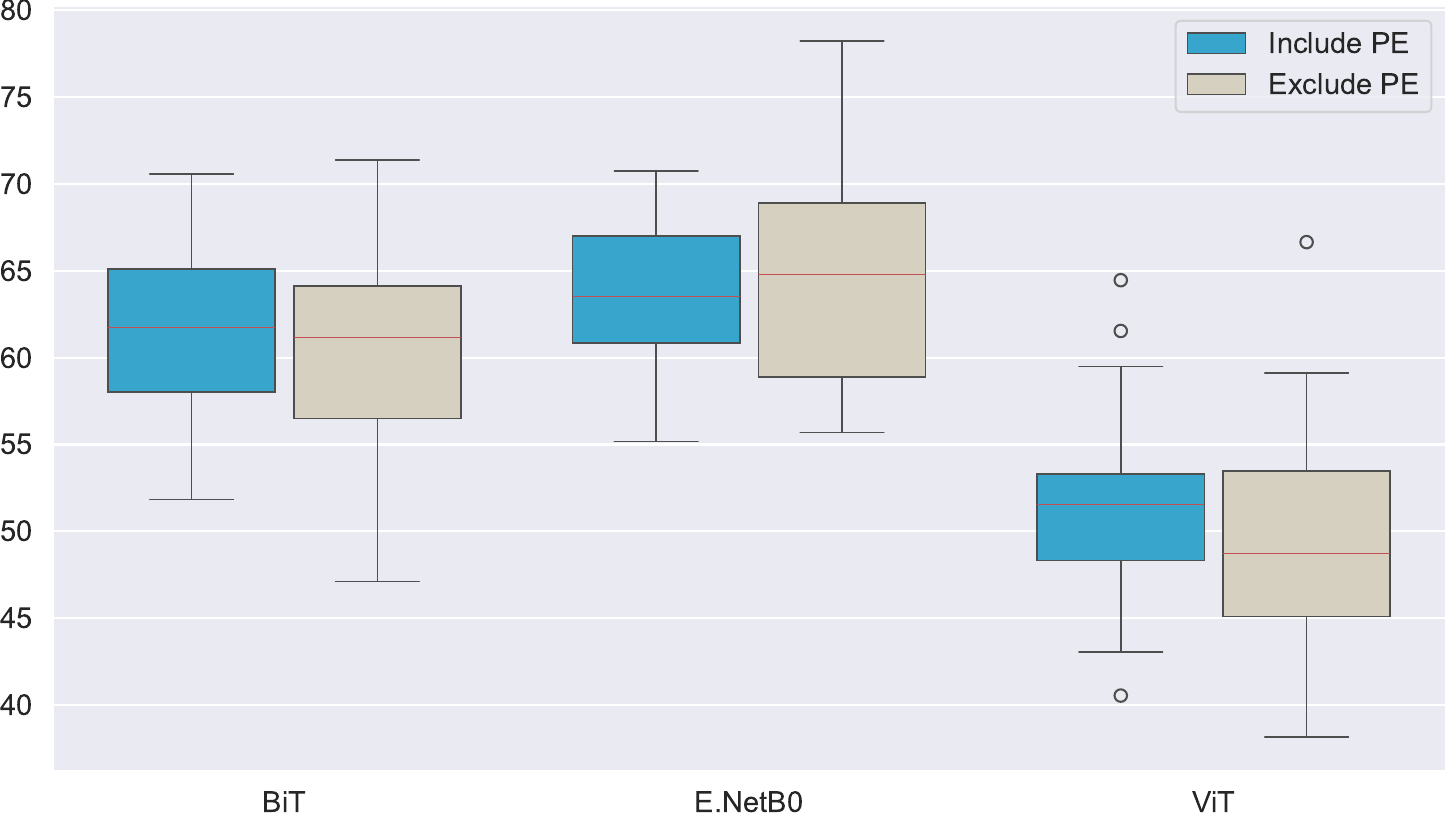}}\par
  \subfloat[][RR]{\includegraphics[width=0.5\columnwidth]{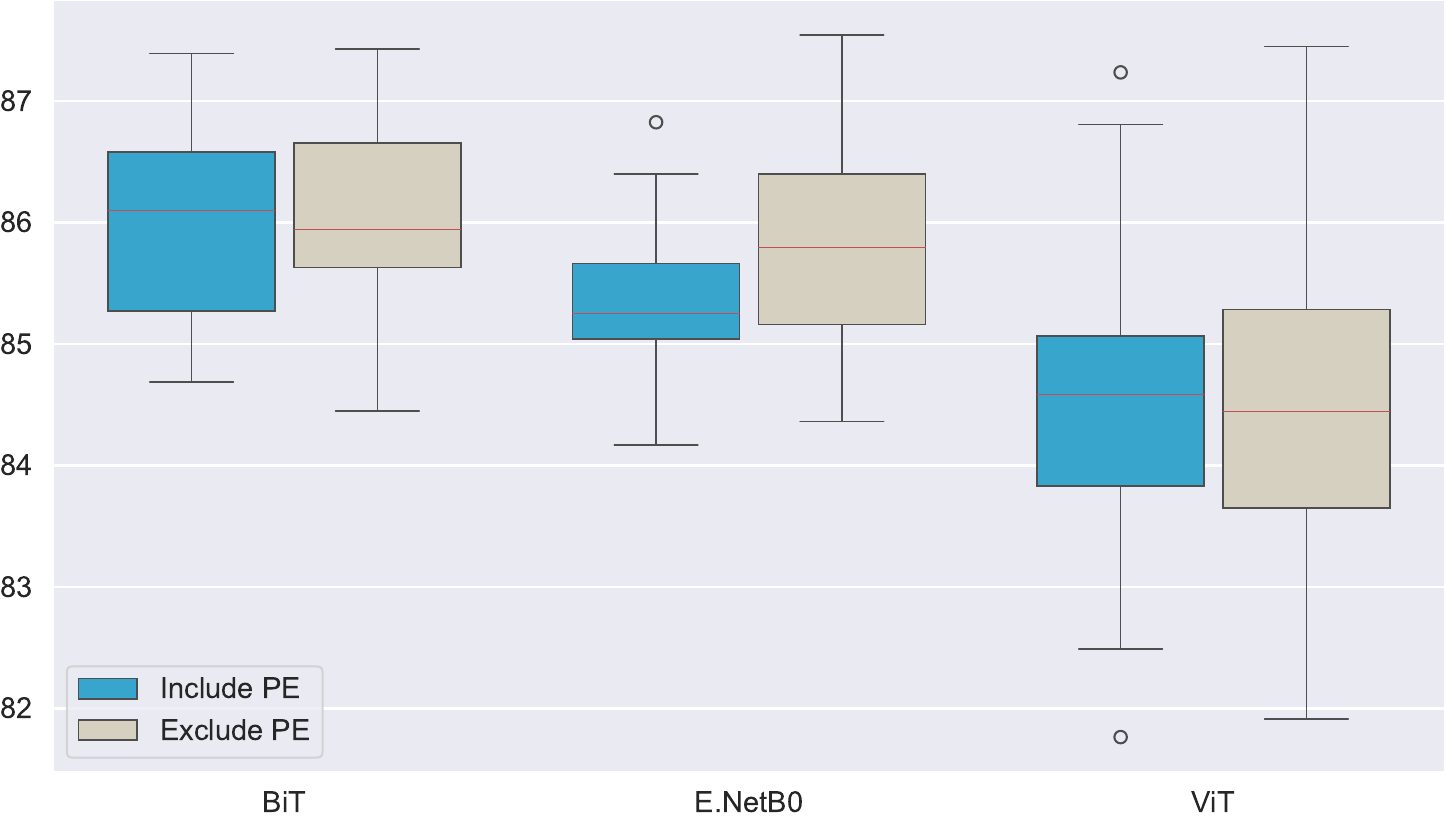}}
  \caption{Comparative Performance of U-SNN With and Without PE as Input at Confidence Threshold 0.1.}
  \label{fig:USNN_performance_comparison_th01}
\end{figure}

To start, let’s first examine the results at confidence threshold 0.1.  Figure \ref{fig:USNN_performance_comparison_th01} illustrates the performance comparison of the U-SNN at  confidence threshold 0.1. At the confidence threshold of 0.1, including PE in the meta-model consistently resulted in superior performance across several key uncertainty metrics when compared to excluding PE. Excluding PE led to a slight increase in CR across all models. Although a higher CR suggests a larger proportion of predictions being labeled as trustworthy, this increase was marginal. The exclusion of PE resulted in a notable deterioration in FCR. For instance, FCR in the ViT model rose from 0.4623 (with PE) to 0.7815 (without PE), marking a 69.1\% increase. Similarly, the BiT model exhibited a 47.4\% increase (from 0.4395 to 0.6478). This sharp increase indicates that excluding PE significantly heightens the risk of confidently incorrect predictions, undermining trust in the model's certainty assessments.

CE also showed a substantial increase when PE was excluded. For example, at confidence threshold 0.1, CE for the ViT model increased by 62.2\% (from 0.9315 to 1.5108), while EfficientNetB0 exhibited a 36.5\% increase. These results underscore PE's critical role in helping the meta-model accurately discern between confidently correct and incorrect predictions.

Excluding PE led to a minor reduction in UR across all models. While this reduction suggests fewer referrals, it also implies that some uncertain and potentially incorrect predictions might be wrongly accepted as confident, contributing to the previously noted increase in FCR and CE. 

To further understand the broader impact of PE, the analysis was extended across all evaluated confidence thresholds. 

Across all thresholds, excluding PE consistently resulted in higher CR values. For example, at 0.4, CR for the ViT model increased from 77.28\% (with PE) to 81.41\% (without PE). Similar trends were observed for BiT and EfficientNetB0. However, while higher CR suggests greater assertiveness in labeling predictions as trustworthy, it does not inherently indicate reliability, especially when juxtaposed with elevated FCR and CE values.

FCR increased notably when PE was excluded, particularly at lower confidence thresholds. For instance, at 0.05, FCR in ViT increased from 0.2212 (with PE) to 0.4078 (without PE), indicating an 84.4\% increase. As thresholds rose, the gap narrowed, but models excluding PE still demonstrated consistently higher FCR values, signaling a persistent vulnerability to confidently incorrect predictions.

The CE metric mirrored FCR trends, with significant increases observed upon PE exclusion. For instance, CE in EfficientNetB0 at 0.2 rose by 30.7\%, from 2.3804 (with PE) to 2.9546 (without PE). The highest differences were again observed at lower thresholds, reaffirming the critical role of PE in reducing confidently incorrect predictions in these ranges.
UR generally decreased when PE was excluded, particularly at higher thresholds. For instance, UR in the ViT model dropped from 22.72\% to 18.59\% at the 0.4 threshold. Although a lower UR could suggest fewer unnecessary referrals, this coincided with higher confidently incorrect predictions, raising concerns about the reliability of certainty assessments without PE. RR remained largely stable across all thresholds and models, regardless of PE inclusion. Minor fluctuations were observed, such as a 1.2\% increase in RR for the BiT model at 0.4, but these differences were not substantial.

\begin{table*}
\centering
\caption{Performance Summary of U-SNN including and excluding PE as input of meta-model}
\label{tab:certainty_result_Exclud_PE}
\resizebox{\linewidth}{!}{%
\begin{tblr}{
  row{1} = {c},
  cell{2}{1} = {r=2}{},
  cell{2}{2} = {r=6}{c},
  cell{4}{1} = {r=2}{},
  cell{6}{1} = {r=2}{},
  cell{8}{1} = {r=2}{},
  cell{8}{2} = {r=6}{c},
  cell{10}{1} = {r=2}{},
  cell{12}{1} = {r=2}{},
  cell{14}{1} = {r=2}{},
  cell{14}{2} = {r=6}{c},
  cell{16}{1} = {r=2}{},
  cell{18}{1} = {r=2}{},
  cell{20}{1} = {r=2}{},
  cell{20}{2} = {r=6}{c},
  cell{22}{1} = {r=2}{},
  cell{24}{1} = {r=2}{},
  hline{1-2,26} = {-}{},
  hline{4,6,10,12,16,18,22,24} = {1,3-8}{dashed,Gray},
  hline{8,14,20} = {-}{Gray},
}
Pretrained & Confidence
  Threshold & U-SNN Input & CR & FCR & CE & UR & RR\\
BiT & 0.05 & Include PE & 29.4514 ± 4.4865 & 0.2081 ± 0.0749 & 0.6902 ± 0.1652 & 70.5486 ± 4.4865 & 87.4771 ± 0.6375\\
 &  & Exclude PE & 29.5792 ± 5.3093 & 0.3093 ± 0.1735 & 0.9804 ± 0.3992 & 70.4208 ± 5.3093 & 87.5615 ± 0.6489\\
E.NetB0 &  & Include PE & 27.4149 ± 3.7112 & 0.2182 ± 0.0897 & 0.7724 ± 0.2322 & 72.5851 ± 3.7112 & 86.7379 ± 0.5236\\
 &  & Exclude PE & 26.4713 ± 5.6186 & 0.3159 ± 0.1889 & 1.1026 ± 0.4884 & 73.5287 ± 5.6186 & 87.0982 ± 0.7385\\
ViT &  & Include PE & 39.3139 ± 5.8126 & 0.2212 ± 0.0981 & 0.5408 ± 0.1706 & 60.6861 ± 5.8126 & 86.4584 ± 1.0562\\
 &  & Exclude PE & 40.988 ± 5.8228 & 0.4078 ± 0.1494 & 0.964 ± 0.2512 & 59.012 ± 5.8228 & 86.3697 ± 1.0404\\
BiT & 0.1 & Include PE & 38.7992 ± 5.0042 & 0.4395 ± 0.1458 & 1.107 ± 0.2486 & 61.2008 ± 5.0042 & 85.9213 ± 0.828\\
 &  & Exclude PE & 39.5851 ± 5.8851 & 0.6478 ± 0.2872 & 1.5713 ± 0.4683 & 60.4149 ± 5.8851 & 86.0389 ± 0.8188\\
E.NetB0 &  & Include PE & 36.1502 ± 3.9059 & 0.4755 ± 0.1488 & 1.2907 ± 0.2857 & 63.8498 ± 3.9059 & 85.3193 ± 0.6299\\
 &  & Exclude PE & 35.4484 ± 6.2704 & 0.6607 ± 0.3294 & 1.7619 ± 0.6362 & 64.5516 ± 6.2704 & 85.8188 ± 0.8694\\
ViT &  & Include PE & 48.3345 ± 5.5188 & 0.4623 ± 0.1625 & 0.9315 ± 0.2372 & 51.6655 ± 5.5188 & 84.5481 ± 1.2343\\
 &  & Exclude PE & 50.4294 ± 5.852 & 0.7815 ± 0.2616 & 1.5108 ± 0.3701 & 49.5706 ± 5.852 & 84.5088 ± 1.2015\\
BiT & 0.2 & Include PE & 52.9509 ± 5.2666 & 1.1213 ± 0.2945 & 2.0862 ± 0.3533 & 47.0491 ± 5.2666 & 83.098 ± 1.1284\\
 &  & Exclude PE & 54.2792 ± 5.4427 & 1.4892 ± 0.4291 & 2.6953 ± 0.5237 & 45.7208 ± 5.4427 & 83.3864 ± 0.9541\\
E.NetB0 &  & Include PE & 50.0759 ± 4.1463 & 1.2038 ± 0.2583 & 2.3804 ± 0.339 & 49.9241 ± 4.1463 & 82.6654 ± 0.8199\\
 &  & Exclude PE & 49.535 ± 6.5867 & 1.5109 ± 0.5621 & 2.9546 ± 0.775 & 50.465 ± 6.5867 & 83.5164 ± 1.1192\\
ViT &  & Include PE & 60.0983 ± 4.7476 & 1.0308 ± 0.2663 & 1.692 ± 0.3233 & 39.9017 ± 4.7476 & 81.4207 ± 1.411\\
 &  & Exclude PE & 63.2945 ± 5.2023 & 1.6386 ± 0.4265 & 2.5508 ± 0.493 & 36.7055 ± 5.2023 & 81.4236 ± 1.3043\\
BiT & 0.4 & Include PE & 75.6858 ± 3.5126 & 3.3056 ± 0.4536 & 4.3513 ± 0.4165 & 24.3142 ± 3.5126 & 76.3087 ± 1.4808\\
 &  & Exclude PE & 78.0664 ± 4.0813 & 4.1313 ± 0.6528 & 5.2644 ± 0.5856 & 21.9336 ± 4.0813 & 77.4471 ± 1.5522\\
E.NetB0 &  & Include PE & 72.9876 ± 3.8683 & 3.4535 ± 0.4571 & 4.7129 ± 0.3947 & 27.0124 ± 3.8683 & 76.2099 ± 1.5066\\
 &  & Exclude PE & 74.436 ± 5.4932 & 4.0678 ± 0.9003 & 5.4068 ± 0.8489 & 25.564 ± 5.4932 & 77.4412 ± 1.7935\\
ViT &  & Include PE & 77.2759 ± 3.3395 & 2.669 ± 0.4377 & 3.4373 ± 0.4289 & 22.7241 ± 3.3395 & 74.6181 ± 1.7885\\
 &  & Exclude PE & 81.4089 ± 3.8916 & 3.7859 ± 0.6173 & 4.6256 ± 0.557 & 18.5911 ± 3.8916 & 74.9084 ± 1.7941
\end{tblr}
}
\end{table*}

\section{Conclusion} \label{section:Conclusion}
The application of AI in high-stakes fields such as medical diagnosis holds immense potential for improving outcomes, yet widespread adoption is hindered by a lack of trust, primarily due to variations in model performance. A key challenge lies in addressing confidently incorrect predictions—cases where the model makes an incorrect prediction with high confidence—alongside the more straightforward detection of uncertain predictions. This research addresses this critical issue by proposing a framework that not only identifies uncertain predictions but also accurately flags confidently incorrect predictions, thereby enhancing trust in ML models. The proposed architecture consists of two levels: a base model (Level 0) that provides predictions and generates a PE measure, and a meta-model (Level 1) that leverages both the feature representations and PE from the base model to learn and predict a trust flag. This trust flag, combined with the predicted class, provides a final output indicating both the class prediction and the associated confidence level, categorized as either trustworthy or untrustworthy.

The evaluation was conducted in two primary stages. First, the performance of U-SNN was compared against the traditional threshold-based method using a series of proposed reliability metrics across multiple confidence thresholds and pre-trained models. The results demonstrated that, at lower confidence thresholds while both approaches produced comparable CR and UR values, the U-SNN framework substantially outperformed the traditional method in reducing confidently incorrect predictions. Specifically, U-SNN consistently exhibited lower FCR and CE values, indicating a stronger ability to prevent confidently incorrect predictions and thereby enhance the reliability of certainty assessments. Moreover, U-SNN achieved lower RR values, suggesting greater efficiency in minimizing unnecessary referrals, which is crucial for reducing operational overhead in real-world scenarios.

A key insight from the comparative analysis was that while the traditional threshold-based method showed competitive performance at higher confidence thresholds (e.g., 0.4), such thresholds may not be practical for high-stakes applications. Lower thresholds, which allow models to be more conservative in assigning certainty, are generally preferable for ensuring reliability. In this context, U-SNN demonstrated a significant advantage, particularly at thresholds below 0.4, where confidently incorrect predictions pose the highest risk.

The second stage of analysis involved an ablation study assessing the impact of including or excluding PE as an input to the meta-model. The findings revealed that including PE consistently enhanced U-SNN's reliability, particularly by reducing FCR and CE values. Although excluding PE led to slightly higher CR values, this came at the cost of increased confidently incorrect predictions, undermining trust in the system's outputs. 

The proposed U-SNN framework provides a robust and reliable alternative to traditional threshold-based uncertainty quantification methods. By leveraging a meta-model and incorporating prediction entropy, U-SNN significantly reduces the risk of confidently incorrect predictions while optimizing the referral process. These advantages are particularly valuable for high-stakes domains, such as healthcare, where trust and decision reliability are paramount. Moreover, the analysis underscores the importance of selecting lower confidence thresholds to enhance model conservativeness and reliability. Future research is directed toward exploring the enhancement of base model calibration to reduce the UR and RR.

% \section*{Acknowledgments}

% {\appendix[Title of appendix]
% }


\bibliographystyle{IEEEtran}
\bibliography{Refs}

\begin{thebibliography}{10}
\providecommand{\url}[1]{#1}
\csname url@samestyle\endcsname
\providecommand{\newblock}{\relax}
\providecommand{\bibinfo}[2]{#2}
\providecommand{\BIBentrySTDinterwordspacing}{\spaceskip=0pt\relax}
\providecommand{\BIBentryALTinterwordstretchfactor}{4}
\providecommand{\BIBentryALTinterwordspacing}{\spaceskip=\fontdimen2\font plus
\BIBentryALTinterwordstretchfactor\fontdimen3\font minus \fontdimen4\font\relax}
\providecommand{\BIBforeignlanguage}[2]{{%
\expandafter\ifx\csname l@#1\endcsname\relax
\typeout{** WARNING: IEEEtran.bst: No hyphenation pattern has been}%
\typeout{** loaded for the language `#1'. Using the pattern for}%
\typeout{** the default language instead.}%
\else
\language=\csname l@#1\endcsname
\fi
#2}}
\providecommand{\BIBdecl}{\relax}
\BIBdecl

\bibitem{jin2020development}
C.~Jin, W.~Chen, Y.~Cao, Z.~Xu, Z.~Tan, X.~Zhang, L.~Deng, C.~Zheng, J.~Zhou, H.~Shi \emph{et~al.}, ``Development and evaluation of an artificial intelligence system for covid-19 diagnosis,'' \emph{Nature communications}, vol.~11, no.~1, p. 5088, 2020.

\bibitem{burger2024unmet}
V.~K. B{\"u}rger, J.~Amann, C.~K. Bui, J.~Fehr, and V.~I. Madai, ``The unmet promise of trustworthy ai in healthcare: why we fail at clinical translation,'' \emph{Frontiers in Digital Health}, vol.~6, p. 1279629, 2024.

\bibitem{cheng2022promoting}
M.~Cheng, X.~Li, and J.~Xu, ``Promoting healthcare workers’ adoption intention of artificial-intelligence-assisted diagnosis and treatment: The chain mediation of social influence and human--computer trust,'' \emph{International Journal of Environmental Research and Public Health}, vol.~19, no.~20, p. 13311, 2022.

\bibitem{fehr2024trustworthy}
J.~Fehr, B.~Citro, R.~Malpani, C.~Lippert, and V.~I. Madai, ``A trustworthy ai reality-check: the lack of transparency of artificial intelligence products in healthcare,'' \emph{Frontiers in Digital Health}, vol.~6, p. 1267290, 2024.

\bibitem{abdar2021review}
M.~Abdar, F.~Pourpanah, S.~Hussain, D.~Rezazadegan, L.~Liu, M.~Ghavamzadeh, P.~Fieguth, X.~Cao, A.~Khosravi, U.~R. Acharya \emph{et~al.}, ``A review of uncertainty quantification in deep learning: Techniques, applications and challenges,'' \emph{Information fusion}, vol.~76, pp. 243--297, 2021.

\bibitem{aseeri2021uncertainty}
A.~O. Aseeri, ``Uncertainty-aware deep learning-based cardiac arrhythmias classification model of electrocardiogram signals,'' \emph{Computers}, vol.~10, no.~6, p.~82, 2021.

\bibitem{gamerman2006markov}
D.~Gamerman and H.~F. Lopes, \emph{Markov chain Monte Carlo: stochastic simulation for Bayesian inference}.\hskip 1em plus 0.5em minus 0.4em\relax Chapman and Hall/CRC, 2006.

\bibitem{graves2011practical}
A.~Graves, ``Practical variational inference for neural networks,'' \emph{Advances in neural information processing systems}, vol.~24, 2011.

\bibitem{gal2016dropout}
Y.~Gal and Z.~Ghahramani, ``Dropout as a bayesian approximation: Representing model uncertainty in deep learning,'' in \emph{international conference on machine learning}.\hskip 1em plus 0.5em minus 0.4em\relax PMLR, 2016, pp. 1050--1059.

\bibitem{kingma2013auto}
D.~P. Kingma and M.~Welling, ``Auto-encoding variational bayes,'' \emph{arXiv preprint arXiv:1312.6114}, 2013.

\bibitem{fortunato2017bayesian}
M.~Fortunato, C.~Blundell, and O.~Vinyals, ``Bayesian recurrent neural networks,'' \emph{arXiv preprint arXiv:1704.02798}, 2017.

\bibitem{martin2024uncertainty}
C.~Mart{\'\i}n~Vicario, D.~Rodr{\'\i}guez~Salas, A.~Maier, S.~Hock, J.~Kuramatsu, B.~Kallmuenzer, F.~Thamm, O.~Taubmann, H.~Ditt, S.~Schwab \emph{et~al.}, ``Uncertainty-aware deep learning for trustworthy prediction of long-term outcome after endovascular thrombectomy,'' \emph{Scientific Reports}, vol.~14, no.~1, p. 5544, 2024.

\bibitem{senousy2021mcua}
Z.~Senousy, M.~M. Abdelsamea, M.~M. Gaber, M.~Abdar, U.~R. Acharya, A.~Khosravi, and S.~Nahavandi, ``Mcua: Multi-level context and uncertainty aware dynamic deep ensemble for breast cancer histology image classification,'' \emph{IEEE Transactions on Biomedical Engineering}, vol.~69, no.~2, pp. 818--829, 2021.

\bibitem{carneiro2020deep}
G.~Carneiro, L.~Z. C.~T. Pu, R.~Singh, and A.~Burt, ``Deep learning uncertainty and confidence calibration for the five-class polyp classification from colonoscopy,'' \emph{Medical image analysis}, vol.~62, p. 101653, 2020.

\bibitem{habibpour2023uncertainty}
M.~Habibpour, H.~Gharoun, M.~Mehdipour, A.~Tajally, H.~Asgharnezhad, A.~Shamsi, A.~Khosravi, and S.~Nahavandi, ``Uncertainty-aware credit card fraud detection using deep learning,'' \emph{Engineering Applications of Artificial Intelligence}, vol. 123, p. 106248, 2023.

\bibitem{westermann2021using}
P.~Westermann and R.~Evins, ``Using bayesian deep learning approaches for uncertainty-aware building energy surrogate models,'' \emph{Energy and AI}, vol.~3, p. 100039, 2021.

\bibitem{yao2024uncertainty}
Y.~Yao, T.~Han, J.~Yu, and M.~Xie, ``Uncertainty-aware deep learning for reliable health monitoring in safety-critical energy systems,'' \emph{Energy}, vol. 291, p. 130419, 2024.

\bibitem{teixeira2023bayesian}
A.~C. Teixeira, H.~Yazdanpanah, A.~Pezente, and M.~Ghassemi, ``Bayesian networks improve out-of-distribution calibration for agribusiness delinquency risk assessment,'' in \emph{Proceedings of the Fourth ACM International Conference on AI in Finance}, 2023, pp. 244--252.

\bibitem{aizpurua2018uncertainty}
J.~I. Aizpurua, V.~M. Catterson, B.~G. Stewart, S.~D. McArthur, B.~Lambert, and J.~G. Cross, ``Uncertainty-aware fusion of probabilistic classifiers for improved transformer diagnostics,'' \emph{IEEE Transactions on Systems, Man, and Cybernetics: Systems}, vol.~51, no.~1, pp. 621--633, 2018.

\bibitem{shamsi2021uncertainty}
A.~Shamsi, H.~Asgharnezhad, S.~S. Jokandan, A.~Khosravi, P.~M. Kebria, D.~Nahavandi, S.~Nahavandi, and D.~Srinivasan, ``An uncertainty-aware transfer learning-based framework for covid-19 diagnosis,'' \emph{IEEE transactions on neural networks and learning systems}, vol.~32, no.~4, pp. 1408--1417, 2021.

\bibitem{habibpour2021uncertainty}
M.~Habibpour, H.~Gharoun, A.~Tajally, A.~Shamsi, H.~Asgharnezhad, A.~Khosravi, and S.~Nahavandi, ``An uncertainty-aware deep learning framework for defect detection in casting products,'' \emph{arXiv preprint arXiv:2107.11643}, 2021.

\bibitem{aguilar2022uncertainty}
E.~Aguilar, B.~Nagarajan, and P.~Radeva, ``Uncertainty-aware selecting for an ensemble of deep food recognition models,'' \emph{Computers in Biology and Medicine}, vol. 146, p. 105645, 2022.

\bibitem{tabarisaadi2022optimized}
P.~Tabarisaadi, A.~Khosravi, S.~Nahavandi, M.~Shafie-Khah, and J.~P. Catal{\~a}o, ``An optimized uncertainty-aware training framework for neural networks,'' \emph{IEEE transactions on neural networks and learning systems}, 2022.

\bibitem{shamsi2023novel}
A.~Shamsi, H.~Asgharnezhad, Z.~Bouchani, K.~Jahanian, M.~Saberi, X.~Wang, I.~Razzak, R.~Alizadehsani, A.~Mohammadi, and H.~Alinejad-Rokny, ``A novel uncertainty-aware deep learning technique with an application on skin cancer diagnosis,'' \emph{Neural Computing and Applications}, vol.~35, no.~30, pp. 22\,179--22\,188, 2023.

\bibitem{shamsi2021uncertainty_loss}
A.~Shamsi, H.~Asgharnezhad, A.~Tajally, S.~Nahavandi, and H.~Leung, ``An uncertainty-aware loss function for training neural networks with calibrated predictions,'' \emph{arXiv preprint arXiv:2110.03260}, 2021.

\bibitem{li2022ultra}
X.~Li, X.~Liang, G.~Luo, W.~Wang, K.~Wang, and S.~Li, ``Ultra: Uncertainty-aware label distribution learning for breast tumor cellularity assessment,'' in \emph{International Conference on Medical Image Computing and Computer-Assisted Intervention}.\hskip 1em plus 0.5em minus 0.4em\relax Springer, 2022, pp. 303--312.

\bibitem{dawood2023uncertainty}
T.~Dawood, C.~Chen, B.~S. Sidhu, B.~Ruijsink, J.~Gould, B.~Porter, M.~K. Elliott, V.~Mehta, C.~A. Rinaldi, E.~Puyol-Ant{\'o}n \emph{et~al.}, ``Uncertainty aware training to improve deep learning model calibration for classification of cardiac mr images,'' \emph{Medical Image Analysis}, vol.~88, p. 102861, 2023.

\bibitem{dawood2024_improving}
T.~Dawood, B.~Ruijsink, R.~Razavi, A.~P. King, and E.~Puyol-Ant{\'o}n, ``Improving deep learning model calibration for cardiac applications using deterministic uncertainty networks and uncertainty-aware training,'' \emph{arXiv preprint arXiv:2405.06487}, 2024.

\bibitem{park2022stacking}
U.~Park, Y.~Kang, H.~Lee, and S.~Yun, ``A stacking heterogeneous ensemble learning method for the prediction of building construction project costs,'' \emph{Applied sciences}, vol.~12, no.~19, p. 9729, 2022.

\bibitem{chatzimparmpas2021empirical}
A.~Chatzimparmpas, R.~M. Martins, K.~Kucher, and A.~Kerren, ``Empirical study: visual analytics for comparing stacking to blending ensemble learning,'' in \emph{2021 23rd International Conference on Control Systems and Computer Science (CSCS)}.\hskip 1em plus 0.5em minus 0.4em\relax IEEE, 2021, pp. 1--8.

\bibitem{wu2023covidx}
Y.~Wu, H.~Gunraj, C.-e.~A. Tai, and A.~Wong, ``Covidx cxr-4: An expanded multi-institutional open-source benchmark dataset for chest x-ray image-based computer-aided covid-19 diagnostics,'' \emph{arXiv preprint arXiv:2311.17677}, 2023.

\bibitem{kaggleCOVIDxCXR4}
``{C}{O}{V}{I}{D}x {C}{X}{R}-4 --- kaggle.com,'' \url{https://www.kaggle.com/datasets/andyczhao/covidx-cxr2/data}, [Accessed 27-07-2024].

\bibitem{tan2019efficientnet}
M.~Tan and Q.~Le, ``Efficientnet: Rethinking model scaling for convolutional neural networks,'' in \emph{International conference on machine learning}.\hskip 1em plus 0.5em minus 0.4em\relax PMLR, 2019, pp. 6105--6114.

\bibitem{kolesnikov2020big}
A.~Kolesnikov, L.~Beyer, X.~Zhai, J.~Puigcerver, J.~Yung, S.~Gelly, and N.~Houlsby, ``Big transfer (bit): General visual representation learning,'' in \emph{Computer Vision--ECCV 2020: 16th European Conference, Glasgow, UK, August 23--28, 2020, Proceedings, Part V 16}.\hskip 1em plus 0.5em minus 0.4em\relax Springer, 2020, pp. 491--507.

\bibitem{dosovitskiy2020image}
A.~Dosovitskiy, L.~Beyer, A.~Kolesnikov, D.~Weissenborn, X.~Zhai, T.~Unterthiner, M.~Dehghani, M.~Minderer, G.~Heigold, S.~Gelly, J.~Uszkoreit, and N.~Houlsby, ``An image is worth 16x16 words: Transformers for image recognition at scale,'' in \emph{International Conference on Learning Representations}, 2021.

\end{thebibliography}

\newpage

\vfill

\end{document}